\documentclass[letterpaper]{article} % DO NOT CHANGE THIS
\usepackage{aaai2026}  % DO NOT CHANGE THIS
\usepackage{times}  % DO NOT CHANGE THIS
\usepackage{helvet}  % DO NOT CHANGE THIS
\usepackage{courier}  % DO NOT CHANGE THIS
\usepackage[hyphens]{url}  % DO NOT CHANGE THIS
\usepackage{graphicx} % DO NOT CHANGE THIS
\usepackage{natbib}  % DO NOT CHANGE THIS AND DO NOT ADD ANY OPTIONS TO IT
\usepackage{caption} % DO NOT CHANGE THIS AND DO NOT ADD ANY OPTIONS TO IT
\usepackage{algorithm}
\usepackage{algorithmic}
\usepackage{multirow}
\usepackage{booktabs}
\usepackage{amsmath}
\usepackage{subcaption}
\usepackage{xcolor}
\usepackage{newfloat}
\usepackage{listings}
\DeclareCaptionStyle{ruled}{labelfont=normalfont,labelsep=colon,strut=off} % DO NOT CHANGE THIS
\floatstyle{ruled}
\newfloat{listing}{tb}{lst}{}
\floatname{listing}{Listing}
\title{Learning to Fast Unrank in Collaborative Filtering Recommendation}
\author{
    Junpeng Zhao\textsuperscript{\rm 1},
    Lin Li\textsuperscript{\rm 1, *},
    Ming Li\textsuperscript{\rm 1,2},
    Amran Bhuiyan\textsuperscript{\rm 2},
    Jimmy Huang\textsuperscript{\rm 2}
}

\affiliations{
    \textsuperscript{\rm 1}School of Computer Science and Artificial Intelligence, Wuhan University of Technology, China\\
    \textsuperscript{\rm 2}Information Retrieval and Knowledge Management Research Laboratory, York University, Canada\\
    \textsuperscript{*}Corresponding author: cathylilin@whut.edu.cn\\
}

\usepackage{bibentry}
\begin{document}

\maketitle

\begin{abstract}
Modern data-driven recommendation systems risk memorizing sensitive user behavioral patterns, raising privacy concerns. Existing recommendation unlearning methods, while capable of removing target data influence, suffer from inefficient unlearning speed and degraded performance, failing to meet real-time unlearning demands. Considering the ranking-oriented nature of recommendation systems, we present unranking, the process of reducing the ranking positions of target items while ensuring the formal guarantees of recommendation unlearning. To achieve efficient unranking, we propose \textit{Learning to Fast Unrank in Collaborative Filtering Recommendation} (L2UnRank), which operates through three key stages: (a) identifying the influenced scope via interaction-based $p$-hop propagation, (b) computing structural and semantic influences for entities within this scope, and (c) performing efficient, ranking-aware parameter updates guided by influence information. Extensive experiments across multiple datasets and backbone models demonstrate L2UnRank's model-agnostic nature, achieving state-of-the-art unranking effectiveness and maintaining recommendation quality comparable to retraining, while also delivering a 50$\times$ speedup over existing methods. Codes are available at \url{https://github.com/Juniper42/L2UnRank}.

\end{abstract}

% Uncomment the following to link to your code, datasets, an extended version or similar.
% You must keep this block between (not within) the abstract and the main body of the paper.
% \begin{links}
%     \link{Code}{https://aaai.org/example/code}
%     \link{Datasets}{https://aaai.org/example/datasets}
%     \link{Extended version}{https://aaai.org/example/extended-version}
% \end{links}

\section{Introduction}

Data-driven recommendation systems are integral to personalizing user experiences across platforms such as e-commerce, social media, and content streaming~\cite{schafer2001commerce}. However, their efficacy hinges on vast amounts of user data, creating significant tension with user privacy~\cite{chen2025privacy}. These models can inadvertently memorize and expose sensitive information~\cite{nguyen2025privacy}, thus necessitating robust privacy-preserving mechanisms~\cite{cheng2024survey}.

Recommendation unlearning offers a promising solution to selectively remove the influence of specific data, particularly user interactions, from trained models~\cite{chen2022recommendation}. Existing recommendation unlearning methods, derived from general machine unlearning methods~\cite{li2024survey}, can be classified two primary categories: (1) \textbf{Exact Unlearning}, represented by \textit{partition-based} methods~\cite{bourtoule2021machine}. These methods suffer from high computational overhead when unlearning targets are distributed across multiple partitions, a challenge exacerbated by the subjective and ad-hoc nature of user unlearning requests in real-world scenarios~\cite{chen2022recommendation}. (2) \textbf{Approximate Unlearning}, primarily comprising \textit{learning-based}~\cite{hao2024general} methods and \textit{influence functions}. The former requires maintaining additional components, which necessitates complex loss function updates for each component when facing frequent unlearning requests, resulting in significant time overhead. The latter achieves efficient unlearning through model parameter adjustments~\cite{zhang2024recommendation}, but it generally treats unlearning targets in isolation and neglects the rich interaction information in collaborative filtering, often resulting in inefficient and low-quality unlearning~\cite{xue2025towards}.

These methods have limitations in real-world recommendation scenarios, as they typically assume infrequent batch processing for unlearning. However, practical recommendation systems must handle continuous and dynamic unlearning requests from large user populations, which demands timely responses~\cite{xu2024machine}. 

\begin{figure*}[t]
    \centering
    \includegraphics[width=0.9\linewidth]{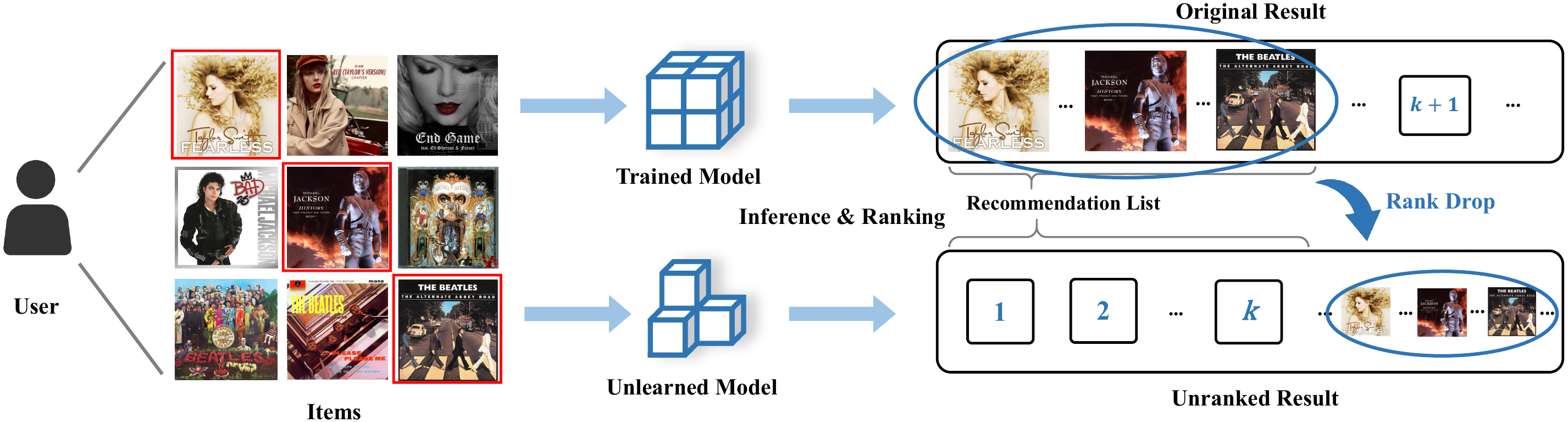}
    \caption{\textbf{An example of unranking.} The red box represents the items to be forgotten.}
    \label{fig:example}
\end{figure*}

To address these limitations, we shift the perspective from data removal to ranking reduction. As illustrated in Figure~\ref{fig:example}, reducing target items' ranking positions achieves the practical objective of recommendation unlearning. When items rank sufficiently low, they fall below users' typical viewing thresholds. Therefore, we propose \textit{unranking}, a paradigm that reformulates recommendation unlearning as a ranking optimization problem, thereby reducing computational complexity and enabling rapid responses to privacy and regulatory requests in dynamic scenarios.

To achieve efficient unranking, we propose \textbf{Learning to Fast Unrank in Collaborative Filtering Recommendation} (L2UnRank), a model-agnostic method that enables efficient ranking-centric unlearning through three synergistic components: (a) \textit{Interaction-based Influence Scoping} identifies influenced entities by exploring the $p$-hop neighborhood of target interactions on the user-item bipartite graph; (b) \textit{Fine-grained Influence Quantification} computes entity popularity and inter-entity similarity to create a smooth, context-aware influence distribution; and (c) \textit{Weighted Influence Function} employs influence-weighted Bayesian Personalized Ranking (BPR)~\cite{rendle2009bpr} loss and conjugate gradient methods to efficiently approximate parameter updates~\cite{takacs2011applications}.

Our method localizes the unlearning process to a small subset of the data, thereby significantly reducing computational overhead. It quantifies the unlearning influence within this scope by leveraging existing information from the recommendation system, such as user and item embeddings. These influence scores are then integrated as weights into the influence function calculation for targeted and efficient parameter adjustments. Our method is validated on three datasets of varying scales and three backbone models of different types. The experimental results confirm its high efficiency, achieving speedups of several orders of magnitude over the retraining baseline.
Our primary contributions are:

\begin{itemize}
    \item [(1)] We introduce \textit{unranking}, a ranking-centric paradigm for recommendation unlearning that aligns the process with the core ranking objective of recommendation systems.
    \item [(2)] We propose L2UnRank, a novel, model-agnostic method that operationalizes unranking through influence scoping, quantification, and efficient approximate updates.
    \item [(3)] Extensive experiments demonstrate L2UnRank's model-agnosticism and superior effectiveness, achieving recommendation quality comparable to retraining with a remarkable 50$\times$ speedup over prior methods.
\end{itemize}

\section{Related work}

\textbf{Machine Unlearning.} Machine unlearning aims to remove the influence of specific data from a trained model, effectively making the model forget what it has learned from that data~\cite{liu2025threats}. This field has gained significant traction due to growing privacy concerns and regulations. The primary challenge lies in achieving this removal efficiently without degrading the model's performance on the remaining data. Broadly, approaches are categorized into \textit{exact unlearning}, which guarantees complete data removal, often at a high computational cost, and \textit{approximate unlearning}, which offers faster alternatives by accepting a small, controlled amount of residual data influence~\cite{li2024survey}.

\textbf{Recommendation Unlearning.} Adapting machine unlearning to recommendation aims to efficiently erase the influence of specific user or interaction data from trained models~\cite{li2024survey}. Current methods fall into two main categories: exact and approximate unlearning. (1) \textbf{Exact unlearning} methods, such as partition-based approaches~\cite{chen2022recommendation}, ensure complete data removal. However, their practical application is hindered by efficiency issues. When unlearning requests are dispersed across numerous data shards, retraining multiple shards becomes necessary, leading to substantial time consumption that fails to meet the high-responsiveness requirements of real-world systems~\cite{li2024survey}. (2) \textbf{Approximate unlearning} provides faster alternatives by tolerating minor residual data influence. Learning-based methods integrate auxiliary modules into recommendation models, but the complex optimization process makes them too slow for real-time or high-frequency unlearning~\cite{li2024towards,hao2024general,liu2022continual}. Influence function-based techniques analytically estimate parameter changes after data deletion~\cite{li2023selective}. Although inherently efficient, they often overlook information within interaction graphs, resulting in extra computation and unstable updates under frequent requests~\cite{zhang2024recommendation}. Another method fine-tunes the model using a reverse ranking objective regularized by the Fisher Information Matrix (FIM)~\cite{you2024rrl}. Its primary bottleneck is the need to recompute the FIM over the entire dataset for each request, which severely undermines its computational efficiency.

In summary, existing recommendation unlearning methods face a trade-off between unlearning quality and efficiency. Exact methods are prohibitively slow for practical applications, while approximate methods, offer improved speed but compromise recommendation utility. Our proposed L2UnRank method is designed to bridge this gap, achieving rapid and effective unlearning while preserving model performance. The technical details are presented in the following section.

\section{Methodology}

\subsection{Problem Formulation}
Let $\mathcal{U}$ and $\mathcal{I}$ denote the sets of users and items, respectively. The set of all user-item interactions is given by $D \subseteq \mathcal{U} \times \mathcal{I}$. A recommendation model $\mathcal{M}$, parameterized by $\Theta$, is trained on $D$ to learn representations for users and items. These representations are then used to predict a preference score $\hat{y}_{ui} = f(\Theta; u, i)$ for a user-item pair $(u, i)$. For each user $u$, items are ranked based on these scores. The rank of item $i$ for user $u$ produced by model $\mathcal{M}$ is $r_{\mathcal{M}}(u, i)$.

\subsubsection{Recommendation Unlearning.}

\begin{figure}[t]
    \centering
    \includegraphics[width=1.0\linewidth]{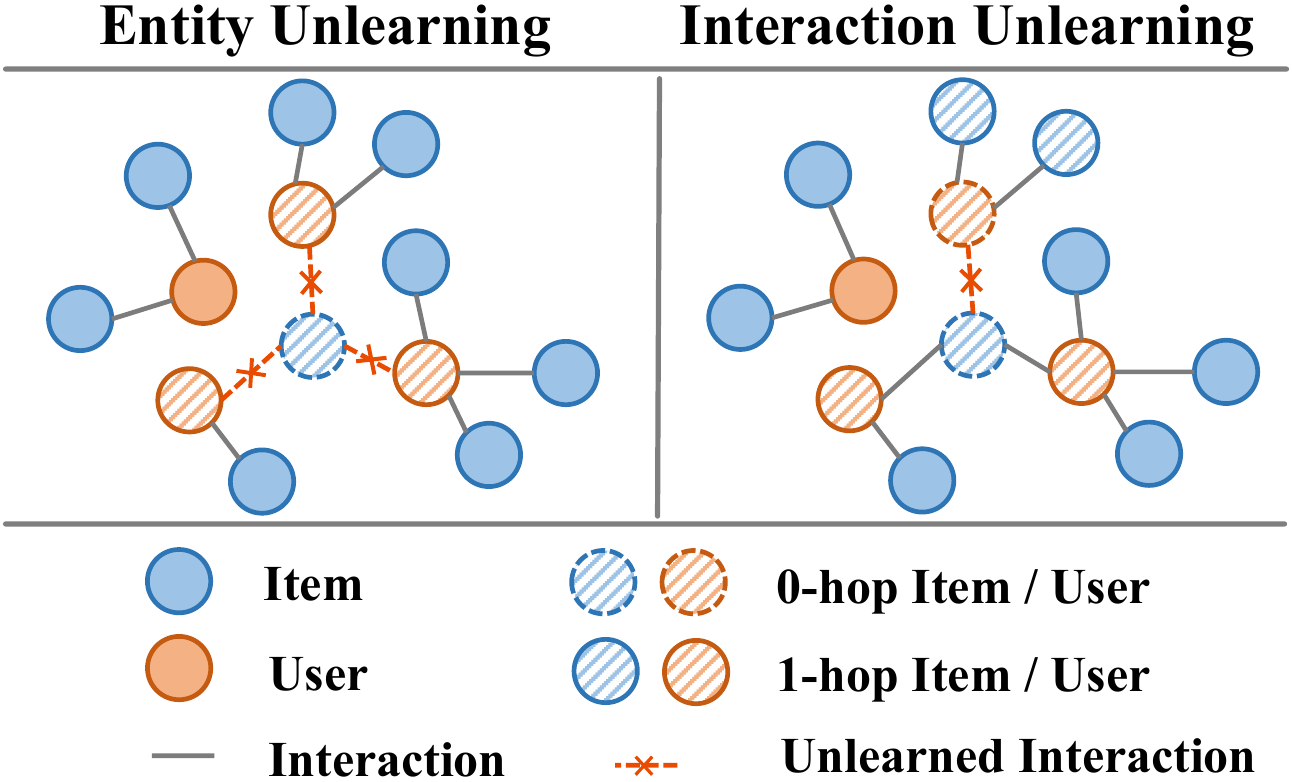}
    \caption{Illustration of recommendation unlearning scenarios. 0-hop items and users refer to directly influenced entities, they and the interactions between them are namely the 0-hop neighborhood.}
    \label{fig:scenarios}
\end{figure}

Current works in recommendation unlearning mainly focus on implicit feedback~\cite{li2024survey}, where the challenge is forgetting interaction. An unlearning task, initiated by a specific request, can be categorized based on its real-world context into two types: \textit{entity unlearning} and \textit{interaction unlearning}. As shown in Figure~\ref{fig:scenarios}, entity unlearning aims to remove all interactions associated with specific users or items, whereas interaction unlearning targets the deletion of specific interactions.

Given an unlearning request, the interaction dataset $D$ is partitioned into a forget set $D_f$ and a retaining set $D_r = D \setminus D_f$. The goal is to generate an unlearned model $\mathcal{M}'$ with parameters $\Theta'$, which can effectively erase the knowledge encoded in $D_f$ while preserving the predictive utility on $D_r$.

\subsubsection{Unranking.}
In scenarios requiring frequent removal of interaction data, we redefine recommendation unlearning as \textit{unranking}. As depicted in Figure~\ref{fig:example}, this method achieves efficient unlearning by perturbing the recommendation list to precisely lower the rank of target items.

In addition to fulfilling the requirements of recommendation unlearning, the unranking must specifically ensure that:
\begin{equation}
    r_{\mathcal{M}'}(u, i) \gg r_{\mathcal{M}}(u, i) \quad \forall (u, i) \in D_f.
\end{equation}
This formulation dictates that for any $(u,i)$ in $\mathcal{D}_f$, its rank must be significantly degraded in the updated model $\mathcal{M}'$.

\subsection{L2UnRank}
\begin{figure*}[t]
    \centering
    \includegraphics[width=1.0\linewidth]{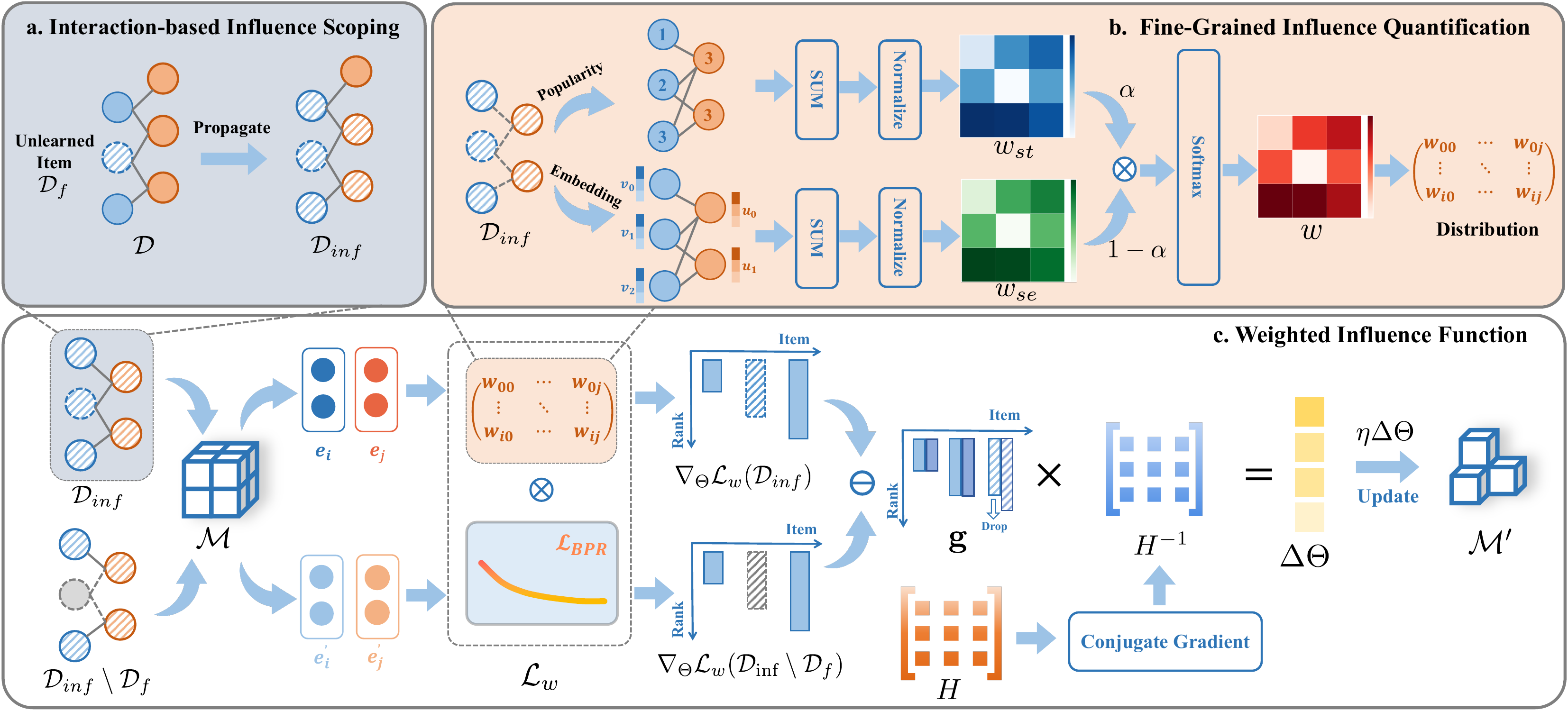}
    \caption{Overview of the proposed L2UnRank method. Deleting an item is taken as an example.}
    \label{fig:overview}
\end{figure*}

To achieve efficient unranking, we propose L2UnRank, a method that effectively approximates the unlearning process without costly retraining. As illustrated in Figure~\ref{fig:overview}, L2UnRank operates in three sequential stages: (1) identifying a localized \textit{Interaction-based Influence Scoping} to constrain the computational cost; (2) performing a \textit{Fine-Grained Influence Quantification} to understand the importance of different entities within this scope; and (3) executing a \textit{Weighted Influence Function} update to achieve fast and precise unranking.

\subsubsection{Interaction-based Influence Scoping.}
To perform unranking efficiently, we must first avoid a full-model update, which is computationally prohibitive. Furthermore, our ultimate goal is to affect item rankings, which requires a loss function that operates on pairwise or listwise comparisons. To address both of these requirements, we begin by identifying a localized subgraph of interactions, termed the \textit{Influenced Scope}, that is most relevant to the forget set $D_f$.

We model the system as a user-item bipartite graph $G = (\mathcal{U} \cup \mathcal{I}, D)$. The influenced scope is constructed by identifying the $p$-hop neighborhood around the entities involved in $D_f$. Let $E_f = \{u, i \mid (u, i) \in D_f\}$ be the set of users and items directly involved in the forget set. We define the 0-hop influenced interactions as $D_{inf}^{(0)} = D_f$. The scope is then expanded iteratively:
\begin{equation}
\label{eq:hop_expansion}
\begin{split}
    D_{inf}^{(k)} = D_{inf}^{(k-1)} \cup \Big\{ (u,i) \in D & \mid \exists (u',i') \in D_{inf}^{(k-1)} \\
    & \text{s.t. } u=u' \lor i=i' \Big\}.
\end{split}
\end{equation}

The final influenced scope is $D_{inf} = D_{inf}^{(p)}$ for a small integer $p$. This process effectively captures the collaborative filtering effect~\cite{koren2021advances}, where interactions involving common entities influence each other. Crucially, this localized set $D_{inf}$ reduces computational complexity while also providing the necessary pairwise data for subsequent ranking updates.

\subsubsection{Fine-Grained Influence Quantification.}

Within the identified scope $D_{inf}$, entities contribute non-uniformly to predictions for $D_f$, with some exerting substantially more influence than others. A precise unranking update thus hinges on identifying and prioritizing these influential entities. We therefore propose to quantify the influence of each entity $v \in E_{inf}$ (where $E_{inf}$ are the entities in $D_{inf}$) by integrating two complementary sources of information: its structural role and its semantic relevance.

\textbf{Structural Influence.} An entity's structural importance is often correlated with its connectivity. We measure this using its degree $d_v$ within the subgraph induced by $D_{inf}$.
\begin{equation}
    w_{st}(v) = d_v, \quad \forall v \in E_{inf}.
\end{equation}

\textbf{Semantic Influence.} To capture content-level relevance, we measure the semantic affinity between an entity $v \in E_{inf}$ and the initial forget entities $E_f$. This is calculated as the aggregated cosine similarity~\cite{xia2015learning} between their embeddings:
\begin{equation}
    w_{se}(v) = \sum_{t \in E_f} \frac{\mathbf{e}_v \cdot \mathbf{e}_t}{\|\mathbf{e}_v\| \|\mathbf{e}_t\|}, \quad \forall v \in E_{inf}.
\end{equation}

\textbf{Unified Influence.}
The structural and semantic scores are first scaled to a consistent range via normalization, denoted by $\mathcal{N}(\cdot)$. These normalized scores are then linearly combined to form a unified raw score:
\begin{equation}
    w_{raw}(v) = \alpha \cdot \mathcal{N}(w_{st}(v)) + (1-\alpha) \cdot \mathcal{N}(w_{se}(v)),
\end{equation}
where $\alpha \in [0, 1]$ is a hyperparameter balancing the two components. Finally, we apply a softmax function~\cite{vaswani2017attention} to obtain a normalized probability distribution:
\begin{equation}
    w(v) = \frac{\exp(w_{raw}(v))}{\sum_{v' \in E_{inf}} \exp(w_{raw}(v'))}, \quad \forall v \in E_{inf}.
\end{equation}
This distribution, $w(v)$, represents the fine-grained influence of each entity, which is pivotal for guiding the subsequent unranking procedure.

\subsubsection{Weighted Influence Function.}
Having identified the relevant interactions and quantified their influence, we now detail the core mechanism for updating the model parameters. Retraining on $D_r$ is infeasible. Instead, we turn to \textit{Influence Functions}~\cite{koh2017understanding}, a technique from robust statistics that can approximate the effect of removing data points on model parameters without retraining.

The standard influence function approximates the parameter change $\Delta\Theta$ as:
\begin{equation}
    \Delta\Theta \approx -H^{-1} \nabla_\Theta \mathcal{L}(D_f) ,
\end{equation}
where $\mathcal{L}$ is the training loss function, $H$ is the Hessian matrix, and $\nabla_\Theta$ denotes the gradient operator with respect to model parameters $\Theta$. 

To adapt this for our unranking task, we must address two key challenges: the choice of loss function and the intractability of the Hessian inverse.

First, to align with our ranking objective, we adopt the Bayesian Personalized Ranking (BPR) loss, which is explicitly designed to optimize the relative order of items. Furthermore, to leverage the influence scores computed previously, we formulate a \textit{weighted} BPR loss that prioritizes training samples based on their computed influence. Specifically, we assign different weights to interaction triplets according to their entity influence scores. For an interaction triplet $(u, i, j)$ consisting of user $u$, positive item $i$, and negative item $j$, the interaction weights are determined by averaging the influence scores of the participating entities.

The weights ensure that interactions involving more influential entities will have a greater impact on the gradient. The weighted BPR loss is:
\begin{equation}
    \mathcal{L}_w(D') = \sum_{(u,i,j) \in D'} -\frac{w(u) + w(i)}{2} \ln \sigma(\hat{y}_{ui} - \hat{y}_{uj}),
\end{equation}
where $D'$ is a set of training triplets and $\sigma$ is the sigmoid function. The gradient contribution of the forget set, $\mathbf{g}$, can then be precisely estimated as the change in the weighted loss gradient over the influenced scope when $D_f$ is removed:
\begin{equation}
    \mathbf{g} = \nabla_{\Theta} \mathcal{L}_w(D_{inf}) - \nabla_{\Theta} \mathcal{L}_w(D_{inf} \setminus D_f).
\end{equation}

Second, to overcome the infeasibility of computing $H^{-1}$, we observe that we only need the Hessian-vector~\cite{chen2020multi} product $H^{-1}\mathbf{g}$, not the inverse itself. This can be obtained by solving the linear system $H\Delta\Theta = -\mathbf{g}$. We employ the \textit{Conjugate Gradient} (CG) ~\cite{nazareth2008conjugate} algorithm to solve this system efficiently. CG is an iterative method that does not require explicitly forming or inverting the Hessian; it only needs a function that can compute the product of the Hessian with an arbitrary vector ($H\mathbf{v}$). This product can be calculated efficiently using automatic differentiation~\cite{pearlmutter1994fast,baydin2018automatic}, making the approach practical for large-scale models.

After obtaining the estimated parameter change $\Delta\Theta$ from CG, the final unlearned parameters are updated as:
\begin{equation}
    \Theta' \leftarrow \Theta + \Delta\Theta / \eta,
\end{equation}
where $\eta \in (0, 1]$ is a scaling factor to control the update step size, ensuring model stability. This targeted update effectively demotes the items in $D_f$ while minimally disturbing the rankings of items in $D_r$.

\begin{table}[t]
    \centering
    \caption{Statistics of the experimental datasets.}
    \label{tab:datasets}
    \small
    \setlength{\tabcolsep}{0.95mm}
    \begin{tabular}{lrrrr}
    \toprule
        \textbf{Dataset} & \textbf{\#Users} & \textbf{\#Items} & \textbf{\#Interactions} & \textbf{Sparsity} \\ \midrule
        MovieLens-1M & 6,040 & 3,706 & 1,000,209 & 95.53\% \\ 
        Yelp2018 & 31,668 & 38,048 & 1,561,406 & 99.87\% \\ 
        Amazon-Book & 52,643 & 91,599 & 2,984,108 & 99.94\% \\ \bottomrule
    \end{tabular}
\end{table}

\subsubsection{Theoretical Analyses.}
The computational complexity of L2UnRank is governed by the CG algorithm, which scales with $O(T \cdot |D_{inf}| \cdot d_{p})$, where $T$ is the number of CG iterations, $|D_{inf}|$ is the size of the influenced scope, and $d_{p}$ is the number of parameters being updated. By localizing the update and $|D_{inf}| \ll |D|$, this is significantly more efficient than full retraining, which has a complexity of $O(E \cdot |D| \cdot d_{p})$ for $E$ epochs. Additionally, we conduct extensive experiments analyzing the convergence properties and numerical stability of the CG algorithm in our proposed method, which are not presented in the main text due to space constraints.

Our method's effectiveness is grounded in the theory of influence functions. Under standard regularity conditions (the loss function is twice-differentiable and the Hessian is positive definite), the approximation error of the influence function is bounded~\cite{koh2017understanding}. The change in the prediction score for a target interaction $(u,i) \in D_f$ is, to a first-order approximation:
\begin{align}
\nonumber s_{\Theta'}(u,i) - s_{\Theta}(u,i) &\approx \nabla_{\Theta} s_{\Theta}(u,i)^T \Delta\Theta \\
&= -\nabla_{\Theta} s_{\Theta}(u,i)^T H^{-1} \mathbf{g}.
\end{align}
This expression shows that the score reduction is maximized when the gradient of the prediction, $\nabla_{\Theta} s_{\Theta}(u,i)$, is aligned with the influence direction $\mathbf{g}$. Our influence quantification and weighted BPR loss are designed to ensure this alignment for interactions truly dependent on the forget set, thereby guaranteeing effective unranking. The use of adaptive scaling factor $\eta$ further ensures that the update does not destabilize the model, thus bounding the potential degradation in utility on the retain set $D_r$.

\begin{table*}[htbp]
    \centering
    \footnotesize
    \caption{Comparison of recommendation accuracy and unlearning time after randomly deleting 5\% of items. Bold indicates the best performance. We report the results for $k=10$, as the performance trends remain consistent for $k=5$ and $k=20$ For recommendation utility, the closer to \textit{Retrain} is better; For unlearning efficiency, the closer to \textit{CertifiedRemoval} is better.}
    \label{tab:accuracy_time}
    \begin{tabular}{@{}llcrcrcr@{}}
        \toprule
        \multirow{2}{*}{\textbf{Backbone}} & \multirow{2}{*}{\textbf{Method}} & \multicolumn{2}{c}{\textbf{MovieLens-1M}} & \multicolumn{2}{c}{\textbf{Yelp2018}} & \multicolumn{2}{c}{\textbf{Amazon-Book}} \\ 
        \cmidrule(lr){3-4} \cmidrule(lr){5-6} \cmidrule(lr){7-8}
        & & \textbf{NDCG@10} & \textbf{Time(s)} & \textbf{NDCG@10} & \textbf{Time(s)} & \textbf{NDCG@10} & \textbf{Time(s)} \\ 
        \midrule
        \multirow{6}{*}{\textbf{LightGCN}} & Retrain & 0.1975 & 47.18 & 0.0394 & 1342.52 & 0.0796 & 8600.60 \\
        & CertifiedRemoval & 0.0328 & 0.54 & 0.0439 & 0.56 & 0.0766 & 1.09 \\
        \cmidrule{2-8}
        & SISA & 0.1023 & 264.24 & 0.0109 & 1757.08 & 0.0261 & 4250.93 \\
        & RecEraser & 0.1994 & 1476.64 & 0.0336 & 2386.66 & 0.0612 & 6236.27 \\
        & IFRU & 0.2020 & 60.57 & 0.0438 & 102.42 & 0.0783 & 199.47 \\
        & Ours & \textbf{0.2024} & \textbf{0.63} & \textbf{0.0440} & \textbf{0.65} & \textbf{0.0785} & \textbf{1.40} \\ 
        \midrule
        \multirow{6}{*}{\textbf{WMF}} & Retrain & 0.1492 & 53.00 & 0.0275 & 115.80 & 0.0180 & 183.50 \\
        & CertifiedRemoval & 0.1048 & 0.47 & 0.0191 & 0.49 & 0.0072 & 0.71 \\
        \cmidrule{2-8}
        & SISA & 0.0745 & 107.80 & 0.0253 & 445.09 & 0.0093 & 723.64 \\
        & RecEraser & 0.0957 & 811.42 & 0.0278 & 2105.77 & 0.0115 & 3442.06 \\
        & IFRU & 0.1532 & 27.14 & 0.0271 & 45.68 & 0.0179 & 87.29 \\
        & Ours & \textbf{0.1578} & \textbf{0.58} & \textbf{0.0296} & \textbf{0.62} & \textbf{0.0235} & \textbf{0.93} \\ 
        \midrule
        \multirow{6}{*}{\textbf{NeuMF}} & Retrain & 0.3094 & 254.56 & 0.0439 & 358.42 & 0.0350 & 1209.88 \\
        & CertifiedRemoval & 0.0338 & 0.42 & 0.0226 & 0.45 & 0.0021 & 0.61 \\
        \cmidrule{2-8}
        & SISA & 0.0669 & 1802.98 & 0.0388 & 3938.32 & 0.0122 & 13669.29 \\
        & RecEraser & 0.3097 & 2390.37 & 0.0430 & 5444.91 & 0.0386 & 15882.28 \\
        & IFRU & 0.3044 & 28.76 & 0.0414 & 44.25 & 0.0398 & 85.02 \\
        & Ours & \textbf{0.3183} & \textbf{0.71} & \textbf{0.0442} & \textbf{0.42} & \textbf{0.0391} & \textbf{1.73} \\ 
        \bottomrule
    \end{tabular}
\end{table*}

\begin{table}[htbp!]
    \centering  \footnotesize
    \caption{Comparison of unlearning effectiveness with LightGCN backbone after randomly deleting 5\% of items. Bold indicates the best performance, underlined indicates the second best performance.}
    \label{tab:unlearning_effectiveness}
    \begin{tabular}{@{}lrcrc@{}}
        \toprule
        \multirow{2}{*}{\textbf{Method}} & \multicolumn{2}{c}{\textbf{MovieLens-1M}} & \multicolumn{2}{c}{\textbf{Yelp2018}} \\ 
        \cmidrule(lr){2-3} \cmidrule(lr){4-5}
        & URR$\uparrow$ & FPR$\uparrow$ & URR$\uparrow$ & FPR$\uparrow$ \\ 
        \midrule
        Retrain          &13.61 &0.0574 &18.96 &0.0137   \\
        CertifiedRemoval &5.39  &0.0825 &15.32 &0.0771   \\
        \cmidrule{2-5}
        SISA             &\underline{5.37}  &0.0261 &\underline{14.00} &0.0107   \\
        RecEraser        &5.13  &\textbf{0.0471} &1.60  &\textbf{0.0313}   \\
        IFRU             &1.22  &0.0210 &0.13  &0.0113   \\
        Ours             &\textbf{23.22} &\underline{0.0458} &\textbf{19.07} &\underline{0.0143}   \\ 
        \bottomrule
    \end{tabular}
\end{table}

\section{Experiments}
Our experiments aim to answer the following research questions (RQs):
\begin{itemize}
    \item \textbf{RQ1:} How does L2UnRank perform compared to existing unlearning baselines across different backbone recommendation models?
    \item \textbf{RQ2:} How do key hyperparameters affect the performance of L2UnRank? % TODO:
    \item \textbf{RQ3:} What is the contribution of each core component in L2UnRank to its overall effectiveness and efficiency?
\end{itemize}

\subsection{Experimental Setup}

\subsubsection{Datasets.}
We conduct experiments on three public benchmark datasets: \textbf{MovieLens-1M}\footnote{https://grouplens.org/datasets/movielens/1m/}, \textbf{Yelp2018}\footnote{https://www.yelp.com/dataset/}, and \textbf{Amazon-Book}\footnote{https://snap.stanford.edu/data/amazon/productGraph/}. These datasets, spanning diverse domains with varying sparsity, are processed by converting all ratings into implicit feedback, following standard practice~\cite{he2020lightgcn, chen2022recommendation}. Table~\ref{tab:datasets} summarizes their statistics.

\subsubsection{Backbones.}
To demonstrate the model-agnostic capability of  L2UnRank, we integrate it with three representative collaborative filtering models:
\begin{itemize}
    \item \textbf{WMF}~\cite{hu2008collaborative}: A classic matrix factorization model optimized for implicit feedback.
    \item \textbf{NeuMF}~\cite{he2017neural}: A seminal deep learning model combining matrix factorization with an MLP to capture non-linear user-item interactions.
    \item \textbf{LightGCN}~\cite{he2020lightgcn}: A graph neural network model that simplifies GCNs for recommendation.
\end{itemize}

\subsubsection{Baselines.}
We compare L2UnRank against five model-agnostic unlearning methods: (1)~\textbf{Retrain}: The gold standard, which retrains a model from scratch on the retained data $D_r$. (2)~\textbf{CertifiedRemoval}~\cite{guo2019certified}: An approximate unlearning method that uses a single Newton-step update based on influence functions; (3)~\textbf{SISA}~\cite{bourtoule2021machine}: A foundational data partitioning method that retrains only the sub-model affected by the data to be forgotten; (4)~\textbf{RecEraser}~\cite{chen2022recommendation}: A SISA-based method for recommendation with balanced data partitioning and adaptive aggregation; and (5)~\textbf{IFRU}~\cite{zhang2024recommendation}: An influence function-based framework for estimating both direct and spillover effects of data removal. 

\subsubsection{Evaluation Metrics.}
Recommendation unlearning methods are typically evaluated from three key perspectives: efficiency, utility, and effectiveness~\cite{fan2025opengu}. NDCG@k and Recall@k are for the evaluation of model utility after unlearning, where k is set to 5, 10 and 20. For unlearning effectiveness, we adopt the \textbf{Unranking Rate} which quantifies ranking degradation of target items~\cite{dang2025efficient}:

\begin{equation}
\text{URR} = \frac{1}{|D_f|} \sum_{(u, i) \in D_f} \frac{r_{(u, i)}' - r_{(u, i)}}{r_{(u, i)} + 1} \mathcal{P},
\end{equation}

Here, $r_{(u,i)}$ and $r'_{(u,i)}$ are the ranks of interaction $(u,i)$ under the original and post-unlearning models, respectively, while $\mathcal{P}$ is the proportion of items in the forget set $D_f$ with a worsened rank. To further assess privacy protection, we also measure the False Positive Rate (FPR) from Membership Inference Attacks (MIA)~\cite{hu2022membership}.

\subsubsection{Implementation Details.}
Following prior work~\cite{wang2019neural}, all models are configured with an embedding size of 64 and a batch size of 1024. They are optimized using the AdamW optimizer~\cite{kingma2014adam} with BPR loss, where the learning rate is selected from \{$10^{-3}, 10^{-4}$\} via grid search. For L2UnRank, the influence scope $p$ is set to 1 for LightGCN and 0 for WMF and NeuMF. Unless specified otherwise, the balancing parameter $\alpha$ and unlearning step size $\eta$ are set to 0.5 and 0.1, respectively. Baselines adhere to their officially recommended configurations. All datasets are partitioned into training, validation, and testing sets at an 8:1:1 ratio. The experiments were conducted on a single NVIDIA A100 GPU, and all reported results are averaged over 10 independent runs. % 遗忘设置，与之前的工作保持一致，2.5%, 5%, 10%表现一致

\begin{figure}[htbp!]
    \centering
    \begin{subfigure}[b]{0.49\columnwidth}
        \includegraphics[width=\columnwidth]{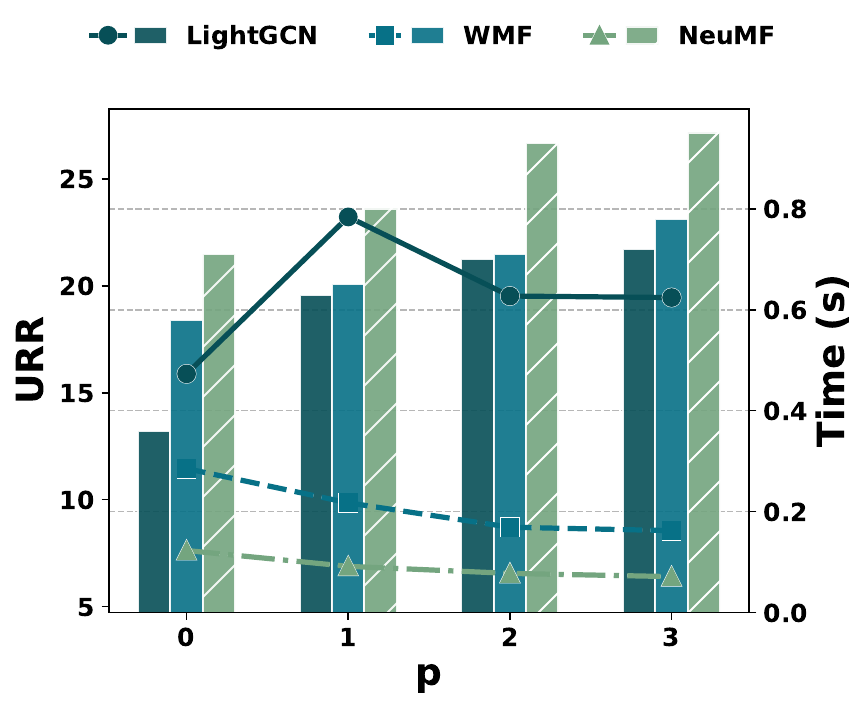}
        \caption{MovieLens-1M.}
        \label{fig:p_ml-1m}
    \end{subfigure}
    \hfill
    \begin{subfigure}[b]{0.49\columnwidth}
        \includegraphics[width=\columnwidth]{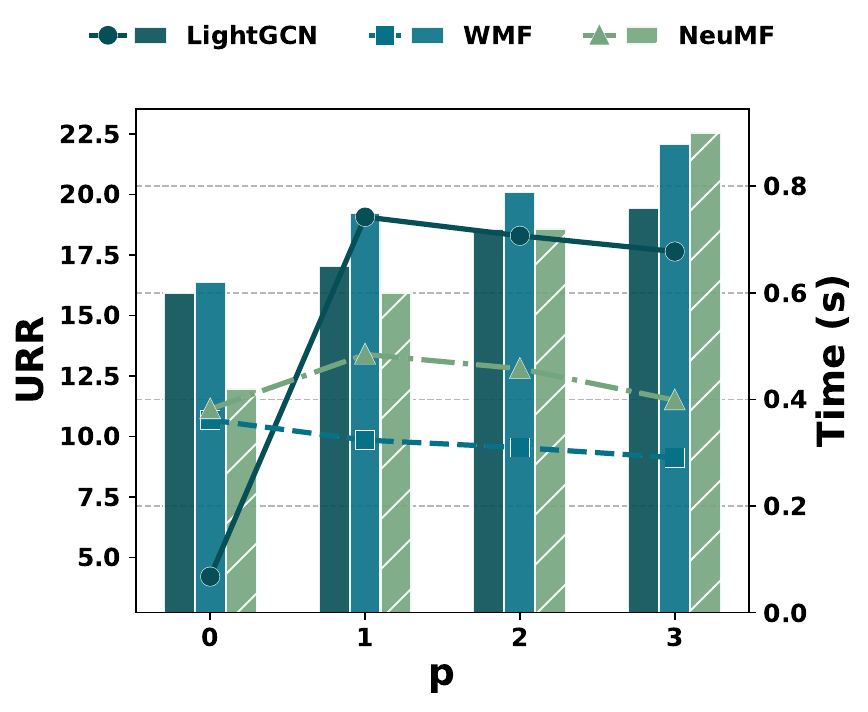}
        \caption{Yelp2018.}
        \label{fig:p_yelp2018}
    \end{subfigure}
    \caption{Effect of influenced scope size $p$ after randomly deleting 5\% of items. The line represents URR, and the bar represents Time (s).}
    \label{fig:influence_scope}
\end{figure}

\begin{figure}[htbp!]
    \centering
    \begin{subfigure}[b]{0.49\columnwidth}
        \includegraphics[width=\columnwidth]{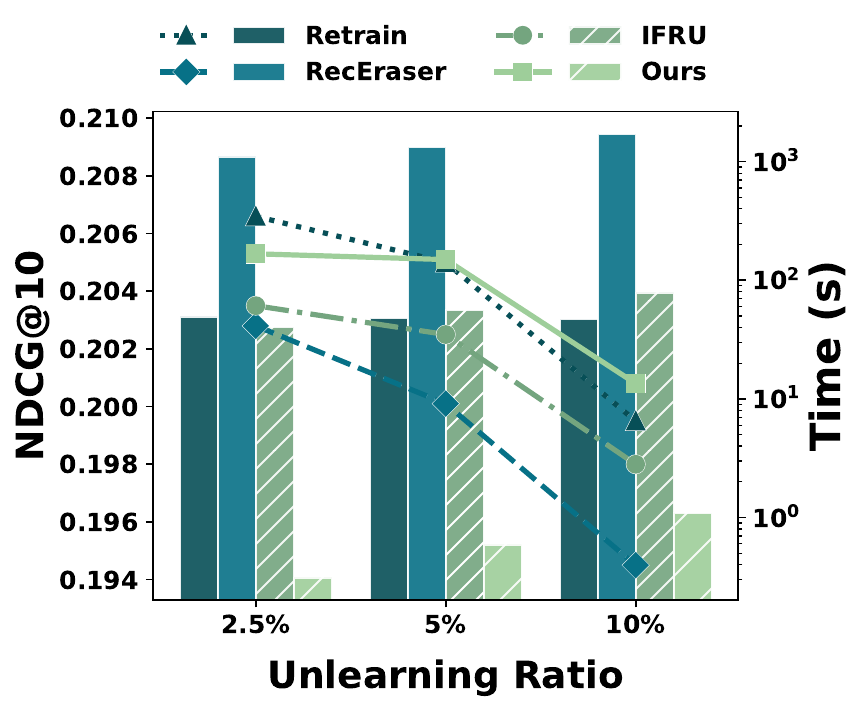}
        \caption{MovieLens-1M.}
        \label{fig:tasks_ml-1m}
    \end{subfigure}
    \hfill
    \begin{subfigure}[b]{0.49\columnwidth}
        \includegraphics[width=\columnwidth]{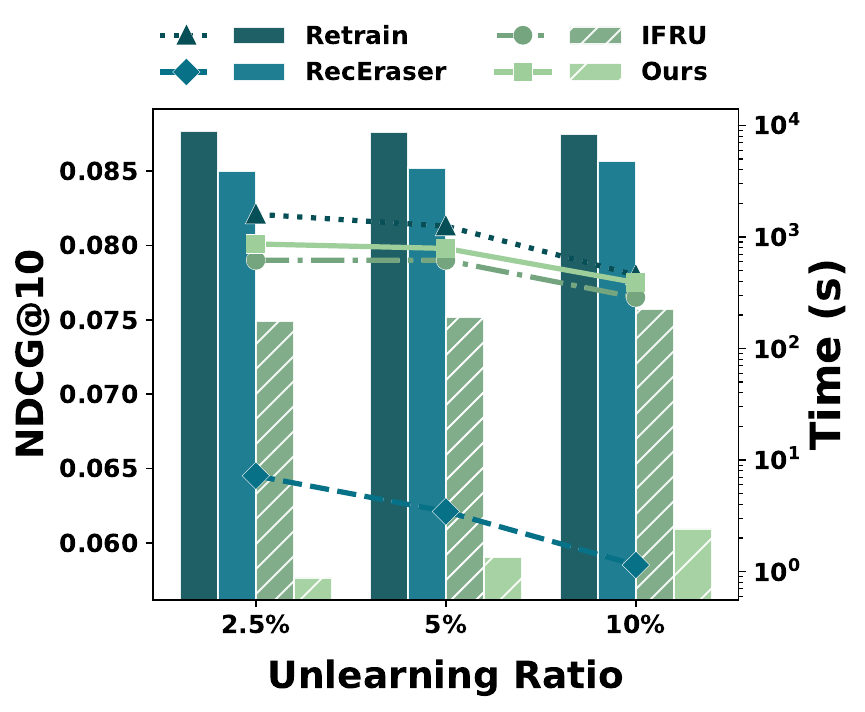}
        \caption{Amazon-Book.}
        \label{fig:tasks_amazon-book}
    \end{subfigure}
    \caption{Robustness evaluation under different Interaction Unlearning ratios.}
    \label{fig:robustness_analysis}
\end{figure}

\subsection{Overall Performance Comparison (RQ1)}
\label{sec_performance_comparison}
Essentially, the goal is to achieve utility close to retraining while maintaining an efficiency similar to certified removal.
\subsubsection{Computational Efficiency.}
As shown in Table~\ref{tab:accuracy_time}, our method is exceptionally efficient, achieving speedups of several orders of magnitude over retraining. For instance, on Amazon-Book with LightGCN, L2UnRank finishes in just 1.4 seconds, whereas retraining takes over 8600 seconds. Our method also significantly outperforms partition-based methods (SISA, RecEraser) by avoiding costly sub-model retraining, and influence-based methods (IFRU) by avoiding the overhead from large, multi-hop neighborhoods. While CertifiedRemoval is fast, its effectiveness is compromised. L2UnRank's localized updates achieve a superior balance between speed and precision.

\subsubsection{Recommendation Utility Preservation.}
L2UnRank excels at preserving model utility. The NDCG@10 results in Table~\ref{tab:accuracy_time} show our method maintains recommendation accuracy competitive with full retraining, sometimes even slightly outperforming it. This phenomenon is likely due to the targeted parameter updates acting as a form of implicit regularization~\cite{wu2024implicit}. In contrast, other approximate methods show significant utility degradation.

\subsubsection{Unlearning Effectiveness.}
Table~\ref{tab:unlearning_effectiveness} shows that L2UnRank consistently achieves the highest URR values, surpassing all baselines, including full retraining. This demonstrates that its weighted influence mechanism is highly effective at targeting parameters responsible for ranking forgotten items, inducing a more significant rank degradation. In the MIA evaluation, L2UnRank achieves a high FPR, even surpassing the SISA. This indicates that our method effectively diminishes the influence of target data within the model, thereby satisfying privacy preservation requirements.
% TODO:
However, we observe a distributional inconsistency between URR and FPR. This suggests that MIA-based metrics alone are insufficient for evaluating the effectiveness of recommendation unlearning~\cite{deeb2024unlearning}. This insufficiency arises because unlearning a specific interaction does not guarantee its removal from the final recommendation list, as rich collaborative information may still lead to its inclusion. The result highlights the need for more comprehensive evaluation metrics that can capture both the direct and indirect effects of recommendation unlearning.

\subsubsection{Robustness Analysis.}
Figure~\ref{fig:robustness_analysis} evaluates L2UnRank's robustness under varying forget ratios. The results show that our method consistently maintains high utility, comparable to retraining, while being significantly faster. Even when forgetting a substantial 10\% of interactions, L2UnRank's performance remains stable, highlighting its ability to handle extensive updates without catastrophic degradation.

\begin{figure}[htbp!]
    \centering
    \begin{subfigure}[b]{0.325\columnwidth}
        \includegraphics[width=\columnwidth]{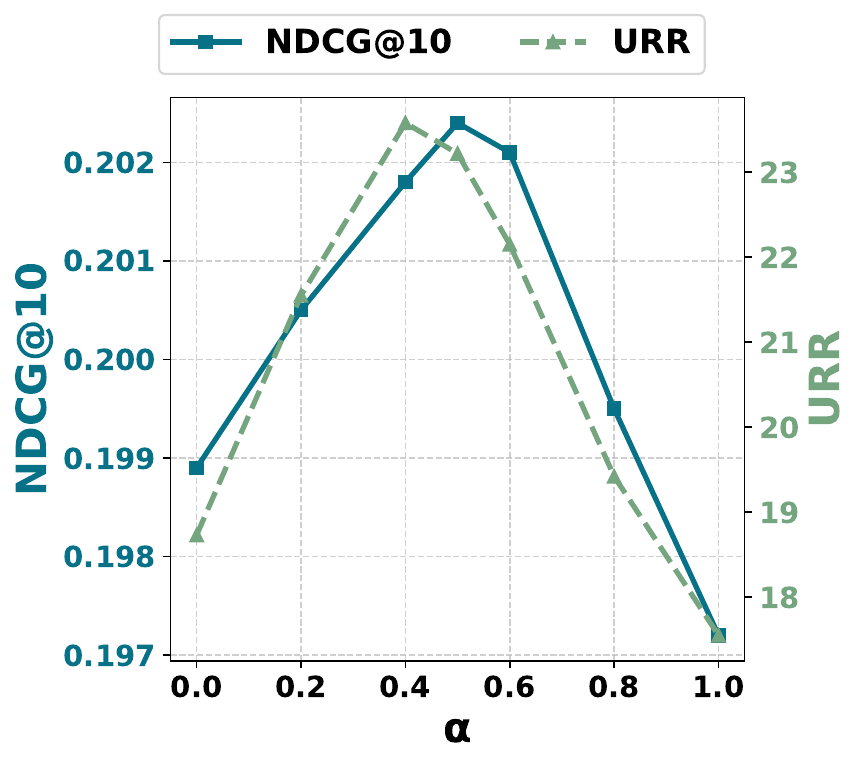}
        \caption{MovieLens-1M.}
        \label{fig:alpha_ml-1m}
    \end{subfigure}
    \hfill
    \begin{subfigure}[b]{0.325\columnwidth}
        \includegraphics[width=\columnwidth]{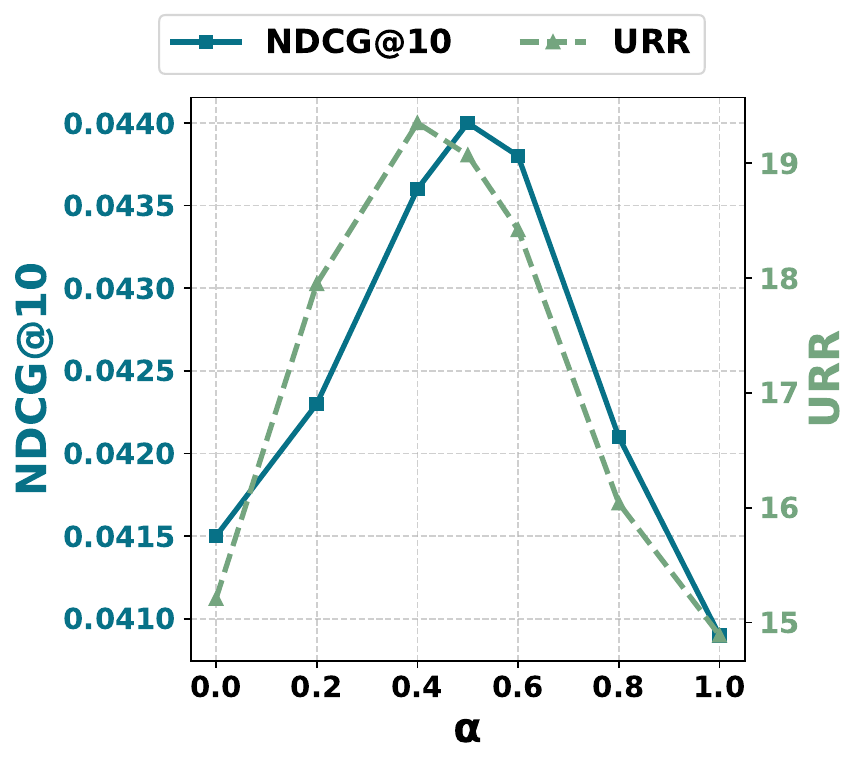}
        \caption{Yelp2018.}
        \label{fig:alpha_yelp2018}
    \end{subfigure}
    \hfill
    \begin{subfigure}[b]{0.325\columnwidth}
        \includegraphics[width=\columnwidth]{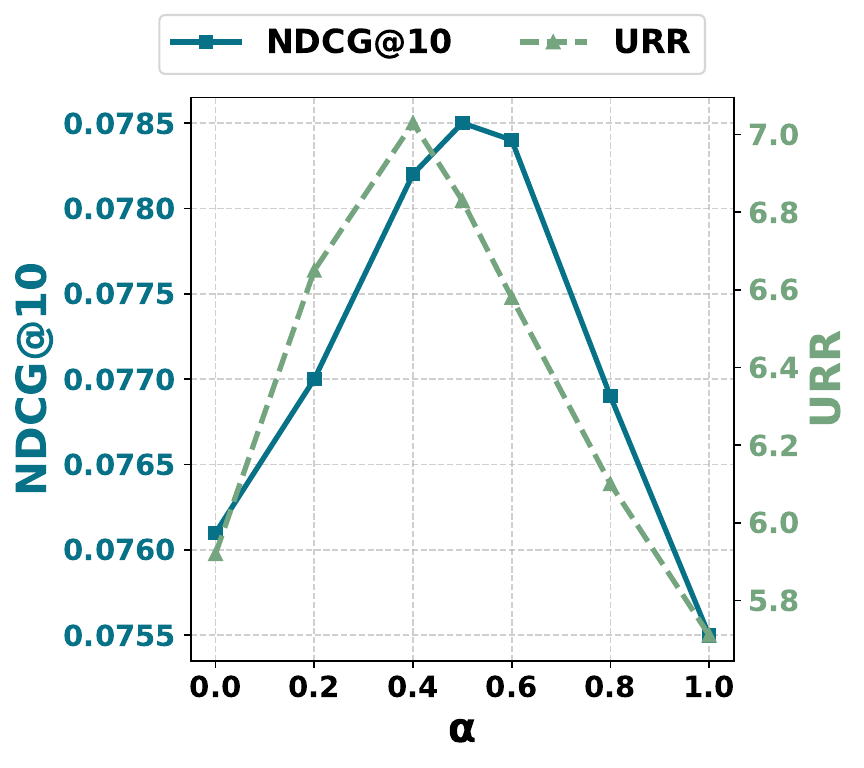}
        \caption{Amazon-Book.}
        \label{fig:alpha-amazon-book}
    \end{subfigure}
    \caption{Impact of influence balancing parameter $\alpha$ using LightGCN across three datasets, after randomly removing 5\% of items.}
    \label{fig:influence_balancing}
\end{figure}

\begin{table}[htbp!]
\centering
\small
\caption{Ablation study results on LightGCN after deleting 5\% of items. ML-1M represents the MovieLens-1M dataset.}
\label{tab:ablation_study}
\footnotesize
\setlength{\tabcolsep}{0.8mm}
\begin{tabular}{llccc}
\toprule
\textbf{Dataset} & \textbf{Variant} & \textbf{Recall@10} & \textbf{URR} & \textbf{Time(s)} \\
\midrule
\multirow{4}{*}{ML-1M}
& w/o Scoping & 0.0691 & 21.55 & 45.8 \\
& w/o Quantification & 0.0665 & 15.68 & 0.61 \\
& w/o RankLoss  & 0.0631 & 3.45  & 0.62 \\
& \textbf{L2UnRank} & \textbf{0.0689} & \textbf{23.22} & \textbf{0.63} \\
\midrule
\multirow{4}{*}{Yelp2018}
& w/o Scoping & 0.0433 & 17.92 & 52.4 \\
& w/o Quantification & 0.0419 & 11.23 & 0.64 \\
& w/o RankLoss  & 0.0390 & 1.88  & 0.65 \\
& \textbf{L2UnRank} & \textbf{0.0431} & \textbf{19.07} & \textbf{0.65} \\
\bottomrule
\end{tabular}
\end{table}

\subsection{Hyperparameter Analysis (RQ2)}
\label{sec_hyperparameter_analysis}

\subsubsection{Impact of Influenced Scope Size $p$.}
As depicted in Figure~\ref{fig:influence_scope}, the optimal influenced scope size $p$ varies across different models. Specifically, LightGCN achieves its best performance at $p=1$, since its message-passing mechanism predominantly captures collaborative signals from 1-hop neighbors. In contrast, WMF and NeuMF, which rely on latent factor interactions rather than graph propagation, perform optimally at $p=0$. This finding is similar with recent research~\cite{ding2025adaptive}.

\subsubsection{Impact of Influence Balancing Parameter $\alpha$.}
Figure~\ref{fig:influence_balancing} shows how $\alpha$ balances structural and semantic influence. Performance peaks when $\alpha$ is in the moderate range of 0.4-0.6, confirming that the two influence types are complementary. Relying solely on one type (i.e., $\alpha=0$ or $\alpha=1$) leads to suboptimal results, as it ignores either network structure or content affinity. By integrating both, L2UnRank creates a balanced influence profile that mitigates issues like the long-tail effect~\cite{anderson2006long} and achieves an optimal trade-off.

\subsection{Component Analysis (RQ3)}
\label{sec_component_analysis}
\subsubsection{Ablation Study.}
Our ablation study, detailed in Table~\ref{tab:ablation_study}, confirms the necessity of each component in L2UnRank. Removing influence scoping (\textit{w/o Scoping}) results in a prohibitive increase in computation time and a diluted unlearning effect, highlighting its importance for both efficiency and precision. Furthermore, replacing fine-grained influence with uniform weights (\textit{w/o Quantification}) yields a poor unranking rate, which demonstrates the criticality of precise influence weighting. Most significantly, substituting the ranking-oriented BPR loss with a classification-based Binary Cross-Entropy (BCE) loss~\cite{ruby2020binary} (\textit{w/o RankLoss}) causes a catastrophic collapse in unlearning performance. This outcome validates our central thesis: the unlearning mechanism must align with the model's primary ranking task, proving that L2UnRank's success stems from the synergy of its components.

\section{Conclusion and Future Work}

In this paper, L2UnRank is introduced, a model-agnostic method that redefines recommendation unlearning as an unranking task. By reducing the ranks of target items, L2UnRank maintains recommendation performance comparable to full retraining while achieving significant unlearning efficiency and effectiveness. Future work will extend this method to sequential models, adapt its influence mechanisms for temporal dynamics, and validate its effectiveness on larger-scale datasets. % TODO: limitation

\bibliography{aaai2026}

@article{li2024survey,
  title={A Survey on Recommendation Unlearning: Fundamentals, Taxonomy, Evaluation, and Open Questions},
  author={Li, Yuyuan and Feng, Xiaohua and Chen, Chaochao and Yang, Qiang},
  journal={arXiv preprint arXiv:2412.12836},
  year={2024}
}

@article{zhang2024recommendation,
  title={Recommendation unlearning via influence function},
  author={Zhang, Yang and Hu, Zhiyu and Bai, Yimeng and Wu, Jiancan and Wang, Qifan and Feng, Fuli},
  journal={ACM Transactions on Recommender Systems},
  volume={3},
  number={2},
  pages={1--23},
  year={2024},
  publisher={ACM New York, NY}
}

@inproceedings{chen2022recommendation,
  title={Recommendation unlearning},
  author={Chen, Chong and Sun, Fei and Zhang, Min and Ding, Bolin},
  booktitle={Proceedings of the ACM web conference 2022},
  pages={2768--2777},
  year={2022}
}

@inproceedings{bourtoule2021machine,
  title={Machine unlearning},
  author={Bourtoule, Lucas and Chandrasekaran, Varun and Choquette-Choo, Christopher A and Jia, Hengrui and Travers, Adelin and Zhang, Baiwu and Lie, David and Papernot, Nicolas},
  booktitle={2021 IEEE symposium on security and privacy (SP)},
  pages={141--159},
  year={2021},
  organization={IEEE}
}

@inproceedings{rendle2009bpr,
  author       = {Steffen Rendle and
                  Christoph Freudenthaler and
                  Zeno Gantner and
                  Lars Schmidt{-}Thieme},
  editor       = {Jeff A. Bilmes and
                  Andrew Y. Ng},
  title        = {{BPR:} Bayesian Personalized Ranking from Implicit Feedback},
  booktitle    = {{UAI} 2009, Proceedings of the Twenty-Fifth Conference on Uncertainty
                  in Artificial Intelligence, Montreal, QC, Canada, June 18-21, 2009},
  pages        = {452--461},
  publisher    = {{AUAI} Press},
  year         = {2009},
  url          = {https://www.auai.org/uai2009/papers/UAI2009\_0139\_48141db02b9f0b02bc7158819ebfa2c7.pdf},
  bibsource    = {dblp computer science bibliography, https://dblp.org}
}

@inproceedings{takacs2011applications,
  title={Applications of the conjugate gradient method for implicit feedback collaborative filtering},
  author={Tak{\'a}cs, G{\'a}bor and Pil{\'a}szy, Istv{\'a}n and Tikk, Domonkos},
  booktitle={Proceedings of the fifth ACM conference on Recommender systems},
  pages={297--300},
  year={2011}
}

@article{chen2025privacy,
  title={Privacy-preserving recommendation with coarse-grained spatiotemporal contexts},
  author={Chen, Lei and Gao, Chen and Lei, Jiahuan and Du, Xiaoyi and Shi, Xinlei and Luo, Hengliang and Jin, Depeng and Li, Yong and Wang, Meng},
  journal={Science China Information Sciences},
  volume={68},
  number={4},
  pages={140104},
  year={2025},
  publisher={Springer}
}

@article{nguyen2025privacy,
  title={Privacy-preserving explainable AI: a survey},
  author={Nguyen, Thanh Tam and Huynh, Thanh Trung and Ren, Zhao and Nguyen, Thanh Toan and Nguyen, Phi Le and Yin, Hongzhi and Nguyen, Quoc Viet Hung},
  journal={Science China Information Sciences},
  volume={68},
  number={1},
  pages={111101},
  year={2025},
  publisher={Springer}
}

@article{cheng2024survey,
  title={A survey on deep neural network pruning: Taxonomy, comparison, analysis, and recommendations},
  author={Cheng, Hongrong and Zhang, Miao and Shi, Javen Qinfeng},
  journal={IEEE Transactions on Pattern Analysis and Machine Intelligence},
  year={2024},
  publisher={IEEE}
}

@inproceedings{hao2024general,
  title={A General Strategy Graph Collaborative Filtering for Recommendation Unlearning},
  author={Hao, Yongjing and Zhuang, Fuzhen and Wang, Deqing and Liu, Guanfeng and Sheng, Victor S and Zhao, Pengpeng},
  booktitle={Proceedings of the 33rd ACM International Conference on Information and Knowledge Management},
  pages={799--808},
  year={2024}
}

@article{liu2025threats,
  title={Threats, attacks, and defenses in machine unlearning: A survey},
  author={Liu, Ziyao and Ye, Huanyi and Chen, Chen and Zheng, Yongsen and Lam, Kwok-Yan},
  journal={IEEE Open Journal of the Computer Society},
  year={2025},
  publisher={IEEE}
}

@article{xu2024machine,
  title={Machine unlearning: Solutions and challenges},
  author={Xu, Jie and Wu, Zihan and Wang, Cong and Jia, Xiaohua},
  journal={IEEE Transactions on Emerging Topics in Computational Intelligence},
  year={2024},
  publisher={IEEE}
}

@article{dang2025efficient,
  title={Efficient and Adaptive Recommendation Unlearning: A Guided Filtering Framework to Erase Outdated Preferences},
  author={Dang, Yizhou and Liu, Yuting and Yang, Enneng and Guo, Guibing and Jiang, Linying and Zhao, Jianzhe and Wang, Xingwei},
  journal={ACM Transactions on Information Systems},
  volume={43},
  number={2},
  pages={1--25},
  year={2025},
  publisher={ACM New York, NY}
}

@inproceedings{cheng2023gnndelete,
  author       = {Jiali Cheng and
                  George Dasoulas and
                  Huan He and
                  Chirag Agarwal and
                  Marinka Zitnik},
  title        = {GNNDelete: {A} General Strategy for Unlearning in Graph Neural Networks},
  booktitle    = {The Eleventh International Conference on Learning Representations,
                  {ICLR} 2023, Kigali, Rwanda, May 1-5, 2023},
  publisher    = {OpenReview.net},
  year         = {2023},
  url          = {https://openreview.net/forum?id=X9yCkmT5Qrl},
  bibsource    = {dblp computer science bibliography, https://dblp.org}
}

@article{pearlmutter1994fast,
  title={Fast exact multiplication by the Hessian},
  author={Pearlmutter, Barak A},
  journal={Neural computation},
  volume={6},
  number={1},
  pages={147--160},
  year={1994},
  publisher={MIT Press}
}

@inproceedings{koh2017understanding,
  title={Understanding black-box predictions via influence functions},
  author={Koh, Pang Wei and Liang, Percy},
  booktitle={International conference on machine learning},
  pages={1885--1894},
  year={2017},
  organization={PMLR}
}

@inproceedings{he2020lightgcn,
  title={Lightgcn: Simplifying and powering graph convolution network for recommendation},
  author={He, Xiangnan and Deng, Kuan and Wang, Xiang and Li, Yan and Zhang, Yongdong and Wang, Meng},
  booktitle={Proceedings of the 43rd International ACM SIGIR conference on research and development in Information Retrieval},
  pages={639--648},
  year={2020}
}

@inproceedings{hu2008collaborative,
  title={Collaborative filtering for implicit feedback datasets},
  author={Hu, Yifan and Koren, Yehuda and Volinsky, Chris},
  booktitle={2008 Eighth IEEE international conference on data mining},
  pages={263--272},
  year={2008},
  organization={Ieee}
}

@inproceedings{he2017neural,
  title={Neural collaborative filtering},
  author={He, Xiangnan and Liao, Lizi and Zhang, Hanwang and Nie, Liqiang and Hu, Xia and Chua, Tat-Seng},
  booktitle={Proceedings of the 26th international conference on world wide web},
  pages={173--182},
  year={2017}
}

@inproceedings{guo2019certified,
  author       = {Chuan Guo and
                  Tom Goldstein and
                  Awni Y. Hannun and
                  Laurens van der Maaten},
  title        = {Certified Data Removal from Machine Learning Models},
  booktitle    = {Proceedings of the 37th International Conference on Machine Learning,
                  {ICML} 2020, 13-18 July 2020, Virtual Event},
  series       = {Proceedings of Machine Learning Research},
  volume       = {119},
  pages        = {3832--3842},
  publisher    = {{PMLR}},
  year         = {2020},
  url          = {http://proceedings.mlr.press/v119/guo20c.html},
  bibsource    = {dblp computer science bibliography, https://dblp.org}
}

@article{li2023selective,
  title={Selective and collaborative influence function for efficient recommendation unlearning},
  author={Li, Yuyuan and Chen, Chaochao and Zheng, Xiaolin and Zhang, Yizhao and Gong, Biao and Wang, Jun and Chen, Linxun},
  journal={Expert Systems with Applications},
  volume={234},
  pages={121025},
  year={2023},
  publisher={Elsevier}
}

@article{fan2025opengu,
  author       = {Bowen Fan and
                  Yuming Ai and
                  Xunkai Li and
                  Zhilin Guo and
                  Rong{-}Hua Li and
                  Guoren Wang},
  title        = {OpenGU: {A} Comprehensive Benchmark for Graph Unlearning},
  journal      = {CoRR},
  volume       = {abs/2501.02728},
  year         = {2025},
  url          = {https://doi.org/10.48550/arXiv.2501.02728},
  doi          = {10.48550/ARXIV.2501.02728},
  eprinttype    = {arXiv},
  eprint       = {2501.02728},
  bibsource    = {dblp computer science bibliography, https://dblp.org}
}

@article{hu2022membership,
  title={Membership inference attacks on machine learning: A survey},
  author={Hu, Hongsheng and Salcic, Zoran and Sun, Lichao and Dobbie, Gillian and Yu, Philip S and Zhang, Xuyun},
  journal={ACM Computing Surveys (CSUR)},
  volume={54},
  number={11s},
  pages={1--37},
  year={2022},
  publisher={ACM New York, NY}
}

@inproceedings{wang2019neural,
  title={Neural graph collaborative filtering},
  author={Wang, Xiang and He, Xiangnan and Wang, Meng and Feng, Fuli and Chua, Tat-Seng},
  booktitle={Proceedings of the 42nd international ACM SIGIR conference on Research and development in Information Retrieval},
  pages={165--174},
  year={2019}
}

@article{ding2025adaptive,
  title={Adaptive Graph Unlearning},
  author={Ding, Pengfei and Wang, Yan and Liu, Guanfeng and Zhu, Jiajie},
  journal={arXiv preprint arXiv:2505.12614},
  year={2025}
}

@misc{anderson2006long,
  title={The long tail},
  author={Anderson, Chris and Nissley, Christopher and Anderson, Chris},
  year={2006},
  publisher={Hyperion Audiobooks}
}

@article{ruby2020binary,
  title={Binary cross entropy with deep learning technique for image classification},
  author={Ruby, Usha and Yendapalli, Vamsidhar and others},
  journal={Int. J. Adv. Trends Comput. Sci. Eng},
  volume={9},
  number={10},
  year={2020}
}

@article{schafer2001commerce,
  title={E-commerce recommendation applications},
  author={Schafer, J Ben and Konstan, Joseph A and Riedl, John},
  journal={Data mining and knowledge discovery},
  volume={5},
  number={1},
  pages={115--153},
  year={2001},
  publisher={Springer}
}

@inproceedings{liu2022continual,
  title={Continual learning and private unlearning},
  author={Liu, Bo and Liu, Qiang and Stone, Peter},
  booktitle={Conference on Lifelong Learning Agents},
  pages={243--254},
  year={2022},
  organization={PMLR}
}

@article{koren2021advances,
  title={Advances in collaborative filtering},
  author={Koren, Yehuda and Rendle, Steffen and Bell, Robert},
  journal={Recommender systems handbook},
  pages={91--142},
  year={2021},
  publisher={Springer}
}

@article{xia2015learning,
  title={Learning similarity with cosine similarity ensemble},
  author={Xia, Peipei and Zhang, Li and Li, Fanzhang},
  journal={Information sciences},
  volume={307},
  pages={39--52},
  year={2015},
  publisher={Elsevier}
}

@article{vaswani2017attention,
  title={Attention is all you need},
  author={Vaswani, Ashish and Shazeer, Noam and Parmar, Niki and Uszkoreit, Jakob and Jones, Llion and Gomez, Aidan N and Kaiser, {\L}ukasz and Polosukhin, Illia},
  journal={Advances in neural information processing systems},
  volume={30},
  year={2017}
}

@article{baydin2018automatic,
  title={Automatic differentiation in machine learning: a survey},
  author={Baydin, Atilim Gunes and Pearlmutter, Barak A and Radul, Alexey Andreyevich and Siskind, Jeffrey Mark},
  journal={Journal of machine learning research},
  volume={18},
  number={153},
  pages={1--43},
  year={2018}
}

@article{wu2024implicit,
  title={Implicit regularization of decentralized gradient descent for sparse regression},
  author={Wu, Tongle and Sun, Ying},
  journal={Advances in Neural Information Processing Systems},
  volume={37},
  pages={16645--16691},
  year={2024}
}

@article{chen2020multi,
  title={Multi-stage influence function},
  author={Chen, Hongge and Si, Si and Li, Yang and Chelba, Ciprian and Kumar, Sanjiv and Boning, Duane and Hsieh, Cho-Jui},
  journal={Advances in Neural Information Processing Systems},
  volume={33},
  pages={12732--12742},
  year={2020}
}

@inproceedings{li2024towards,
  author       = {Xunkai Li and
                  Yulin Zhao and
                  Zhengyu Wu and
                  Wentao Zhang and
                  Rong{-}Hua Li and
                  Guoren Wang},
  editor       = {Michael J. Wooldridge and
                  Jennifer G. Dy and
                  Sriraam Natarajan},
  title        = {Towards Effective and General Graph Unlearning via Mutual Evolution},
  booktitle    = {Thirty-Eighth {AAAI} Conference on Artificial Intelligence, {AAAI}
                  2024, Thirty-Sixth Conference on Innovative Applications of Artificial
                  Intelligence, {IAAI} 2024, Fourteenth Symposium on Educational Advances
                  in Artificial Intelligence, {EAAI} 2014, February 20-27, 2024, Vancouver,
                  Canada},
  pages        = {13682--13690},
  publisher    = {{AAAI} Press},
  year         = {2024},
  url          = {https://doi.org/10.1609/aaai.v38i12.29273},
  doi          = {10.1609/AAAI.V38I12.29273},
  bibsource    = {dblp computer science bibliography, https://dblp.org}
}

@incollection{nazareth2008conjugate,
  title={Conjugate-gradient Methods},
  author={Nazareth, John L},
  booktitle={Encyclopedia of Optimization},
  pages={466--470},
  year={2008},
  publisher={Springer}
}

@article{deeb2024unlearning,
  title={Do unlearning methods remove information from language model weights?},
  author={Deeb, Aghyad and Roger, Fabien},
  journal={arXiv preprint arXiv:2410.08827},
  year={2024}
}

@inproceedings{kingma2014adam,
  author       = {Diederik P. Kingma and
                  Jimmy Ba},
  editor       = {Yoshua Bengio and
                  Yann LeCun},
  title        = {Adam: {A} Method for Stochastic Optimization},
  booktitle    = {3rd International Conference on Learning Representations, {ICLR} 2015,
                  San Diego, CA, USA, May 7-9, 2015, Conference Track Proceedings},
  year         = {2015},
  url          = {http://arxiv.org/abs/1412.6980},
  bibsource    = {dblp computer science bibliography, https://dblp.org}
}

@inproceedings{you2024rrl,
  author       = {Xiaoyu You and
                  Jianwei Xu and
                  Mi Zhang and
                  Zechen Gao and
                  Min Yang},
  editor       = {Michael J. Wooldridge and
                  Jennifer G. Dy and
                  Sriraam Natarajan},
  title        = {{RRL:} Recommendation Reverse Learning},
  booktitle    = {Thirty-Eighth {AAAI} Conference on Artificial Intelligence, {AAAI}
                  2024, Thirty-Sixth Conference on Innovative Applications of Artificial
                  Intelligence, {IAAI} 2024, Fourteenth Symposium on Educational Advances
                  in Artificial Intelligence, {EAAI} 2014, February 20-27, 2024, Vancouver,
                  Canada},
  pages        = {9296--9304},
  publisher    = {{AAAI} Press},
  year         = {2024},
  url          = {https://doi.org/10.1609/aaai.v38i8.28782},
  doi          = {10.1609/AAAI.V38I8.28782},
  bibsource    = {dblp computer science bibliography, https://dblp.org}
}

@article{xue2025towards,
  title={Towards Reliable Forgetting: A Survey on Machine Unlearning Verification, Challenges, and Future Directions},
  author={Xue, Lulu and Hu, Shengshan and Lu, Wei and Shen, Yan and Li, Dongxu and Guo, Peijin and Zhou, Ziqi and Li, Minghui and Zhang, Yanjun and Zhang, Leo Yu},
  journal={arXiv preprint arXiv:2506.15115},
  year={2025}
}
\newpage
\clearpage
\setlength{\leftmargini}{20pt}
\makeatletter\def\@listi{\leftmargin\leftmargini \topsep .5em \parsep .5em \itemsep .5em}
\def\@listii{\leftmargin\leftmarginii \labelwidth\leftmarginii \advance\labelwidth-\labelsep \topsep .4em \parsep .4em \itemsep .4em}
\def\@listiii{\leftmargin\leftmarginiii \labelwidth\leftmarginiii \advance\labelwidth-\labelsep \topsep .4em \parsep .4em \itemsep .4em}\makeatother

\setcounter{secnumdepth}{0}
\renewcommand\thesubsection{\arabic{subsection}}
\renewcommand\labelenumi{\thesubsection.\arabic{enumi}}

\newcounter{checksubsection}
\newcounter{checkitem}[checksubsection]

\newcommand{\checksubsection}[1]{%
  \refstepcounter{checksubsection}%
  \paragraph{\arabic{checksubsection}. #1}%
  \setcounter{checkitem}{0}%
}

\newcommand{\checkitem}{%
  \refstepcounter{checkitem}%
  \item[\arabic{checksubsection}.\arabic{checkitem}.]%
}
\newcommand{\question}[2]{\normalcolor\checkitem #1 #2 \color{blue}}
\newcommand{\ifyespoints}[1]{\makebox[0pt][l]{\hspace{-15pt}\normalcolor #1}}

\section*{Reproducibility Checklist}

\vspace{1em}
\hrule
\vspace{1em}

% The questions start here

\checksubsection{General Paper Structure}
\begin{itemize}

\question{Includes a conceptual outline and/or pseudocode description of AI methods introduced}{(yes/partial/no/NA)}
yes

\question{Clearly delineates statements that are opinions, hypothesis, and speculation from objective facts and results}{(yes/no)}
yes

\question{Provides well-marked pedagogical references for less-familiar readers to gain background necessary to replicate the paper}{(yes/no)}
yes

\end{itemize}
\checksubsection{Theoretical Contributions}
\begin{itemize}

\question{Does this paper make theoretical contributions?}{(yes/no)}
yes

	\ifyespoints{\vspace{1.2em}If yes, please address the following points:}
        \begin{itemize}
	
	\question{All assumptions and restrictions are stated clearly and formally}{(yes/partial/no)}
	yes

	\question{All novel claims are stated formally (e.g., in theorem statements)}{(yes/partial/no)}
	yes

	\question{Proofs of all novel claims are included}{(yes/partial/no)}
	yes

	\question{Proof sketches or intuitions are given for complex and/or novel results}{(yes/partial/no)}
	yes

	\question{Appropriate citations to theoretical tools used are given}{(yes/partial/no)}
	yes

	\question{All theoretical claims are demonstrated empirically to hold}{(yes/partial/no/NA)}
	yes

	\question{All experimental code used to eliminate or disprove claims is included}{(yes/no/NA)}
	yes
	
	\end{itemize}
\end{itemize}

\checksubsection{Dataset Usage}
\begin{itemize}

\question{Does this paper rely on one or more datasets?}{(yes/no)}
yes

\ifyespoints{If yes, please address the following points:}
\begin{itemize}

	\question{A motivation is given for why the experiments are conducted on the selected datasets}{(yes/partial/no/NA)}
	yes

	\question{All novel datasets introduced in this paper are included in a data appendix}{(yes/partial/no/NA)}
	NA

	\question{All novel datasets introduced in this paper will be made publicly available upon publication of the paper with a license that allows free usage for research purposes}{(yes/partial/no/NA)}
	NA

	\question{All datasets drawn from the existing literature (potentially including authors' own previously published work) are accompanied by appropriate citations}{(yes/no/NA)}
	yes

	\question{All datasets drawn from the existing literature (potentially including authors' own previously published work) are publicly available}{(yes/partial/no/NA)}
	yes

	\question{All datasets that are not publicly available are described in detail, with explanation why publicly available alternatives are not scientifically satisficing}{(yes/partial/no/NA)}
	NA

\end{itemize}
\end{itemize}

\checksubsection{Computational Experiments}
\begin{itemize}

\question{Does this paper include computational experiments?}{(yes/no)}
yes

\ifyespoints{If yes, please address the following points:}
\begin{itemize}

	\question{This paper states the number and range of values tried per (hyper-) parameter during development of the paper, along with the criterion used for selecting the final parameter setting}{(yes/partial/no/NA)}
	yes

	\question{Any code required for pre-processing data is included in the appendix}{(yes/partial/no)}
	yes

	\question{All source code required for conducting and analyzing the experiments is included in a code appendix}{(yes/partial/no)}
	yes

	\question{All source code required for conducting and analyzing the experiments will be made publicly available upon publication of the paper with a license that allows free usage for research purposes}{(yes/partial/no)}
	yes
        
	\question{All source code implementing new methods have comments detailing the implementation, with references to the paper where each step comes from}{(yes/partial/no)}
	yes

	\question{If an algorithm depends on randomness, then the method used for setting seeds is described in a way sufficient to allow replication of results}{(yes/partial/no/NA)}
	yes

	\question{This paper specifies the computing infrastructure used for running experiments (hardware and software), including GPU/CPU models; amount of memory; operating system; names and versions of relevant software libraries and frameworks}{(yes/partial/no)}
	yes

	\question{This paper formally describes evaluation metrics used and explains the motivation for choosing these metrics}{(yes/partial/no)}
	yes

	\question{This paper states the number of algorithm runs used to compute each reported result}{(yes/no)}
	yes

	\question{Analysis of experiments goes beyond single-dimensional summaries of performance (e.g., average; median) to include measures of variation, confidence, or other distributional information}{(yes/no)}
	yes

	\question{The significance of any improvement or decrease in performance is judged using appropriate statistical tests (e.g., Wilcoxon signed-rank)}{(yes/partial/no)}
	yes

	\question{This paper lists all final (hyper-)parameters used for each model/algorithm in the paper’s experiments}{(yes/partial/no/NA)}
	yes

\end{itemize}
\end{itemize}

% ---

\appendix
\onecolumn
\section{Experiments Setup}

\subsection{Experimental Configuration}
\label{app_experiments_setup}

To ensure fair comparison across all methods, we adopt a unified experimental configuration with the following settings:

\textbf{Model Configuration.} All models employ an embedding dimension of 64 and a batch size of 1024~\cite{wang2019neural}, optimized using the AdamW optimizer~\cite{kingma2014adam}. The models are trained with BPR loss using a learning rate selected from $\{10^{-3}, 10^{-4}\}$ via grid search.

\textbf{L2UnRank Hyperparameters.} We configure the Influenced scope parameter as $p=1$ for LightGCN (to capture graph-based collaborative filtering effects) and $p=0$ for WMF and NeuMF (due to their latent factor-based nature). The influence balancing factor is set to $\alpha=0.5$ and the scaling factor to $\eta=0.1$ (Appendix~\ref{app_eta}) unless otherwise specified.

\textbf{Baseline Configurations.} All baseline methods are configured following their original papers: \textbf{SISA} uses 10 shards with uniform aggregation strategy; \textbf{RecEraser} adopts the Interaction-based Balanced Partition algorithm with 10 shards and attention size $k=32$, L$_2$ regularization $\lambda=1e^{-4}$; \textbf{CertifiedRemoval} sets L$_2$ regularization $\lambda=1e^{-4}$, target perturbation standard deviation $\sigma=10.0$, and optimization steps to 50; \textbf{IFRU} uses neighbor order $k=1$ with pruning ratios $(a_0, a_1) = (1.0, 1.0)$ for matrix factorization models and $(a_0, a_1) = (1.0, 0.6)$ for LightGCN.

\textbf{Data Partitioning.} Each dataset is partitioned using an 8:1:1 split for training, validation, and testing, respectively.

\textbf{Statistical Reliability.} To ensure statistical reliability, we conduct 10 independent runs with different random seeds and report the averaged results.

\textbf{Hardware Environment.} All experiments are conducted on a machine equipped with an Intel(R) Xeon(R) Platinum 8468 CPU, 120GB RAM, and a single NVIDIA A100 GPU.

We evaluate L2UnRank under two distinct unlearning scenarios:
\begin{itemize}
    \item [(1)] \textbf{Entity Unlearning}: randomly removing [2.5\%, 5\%, 10\%] of items from the system;
    \item [(2)] \textbf{Interaction Unlearning}: randomly selecting [2.5\%, 5\%, 10\%] of users and removing half of their historical interactions.
\end{itemize}

\subsection{Membership Inference Attack}

We implement Membership Inference Attack (MIA) as a privacy evaluation metric to assess the unlearning effectiveness. We employ a 4-layer MLP classifier with ReLU activations, BatchNorm, and Dropout (0.3) for regularization. The member set consists of interactions from the forget set $D_f$, while the non-member set contains randomly generated user-item pairs not in the original training set. We adopt a pure black-box attack approach that only accesses model prediction outputs without requiring internal parameters or shadow models. Features are constructed based on prediction score differences between the original and unlearned models. The FPR metric measures the proportion of non-member samples incorrectly classified as members, serving as a key privacy preservation indicator.

\section{Scaling Factor Analysis}
\label{app_eta}

We conduct a comprehensive analysis of the scaling factor $\eta$ to understand its impact on both recommendation quality and unlearning effectiveness across different datasets and backbone models. The scaling factor controls the magnitude of parameter updates in our influence function-based approach, directly affecting the trade-off between unlearning effectiveness and model utility preservation.

\begin{figure}[ht]
    \centering
    \begin{subfigure}[b]{0.33\columnwidth}
        \includegraphics[width=\columnwidth]{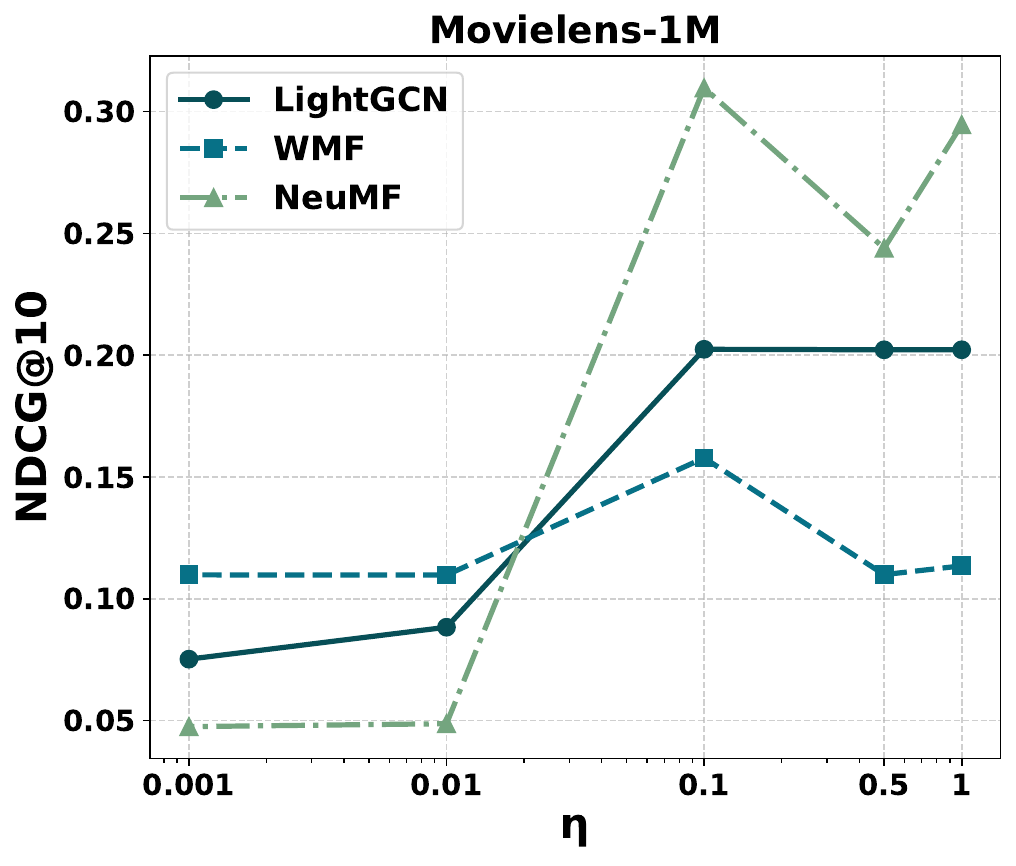}
        % \caption{MovieLens-1M}
        \label{fig:eta_ndcg_ml-1m}
    \end{subfigure}
    \hfill
    \begin{subfigure}[b]{0.33\columnwidth}
        \includegraphics[width=\columnwidth]{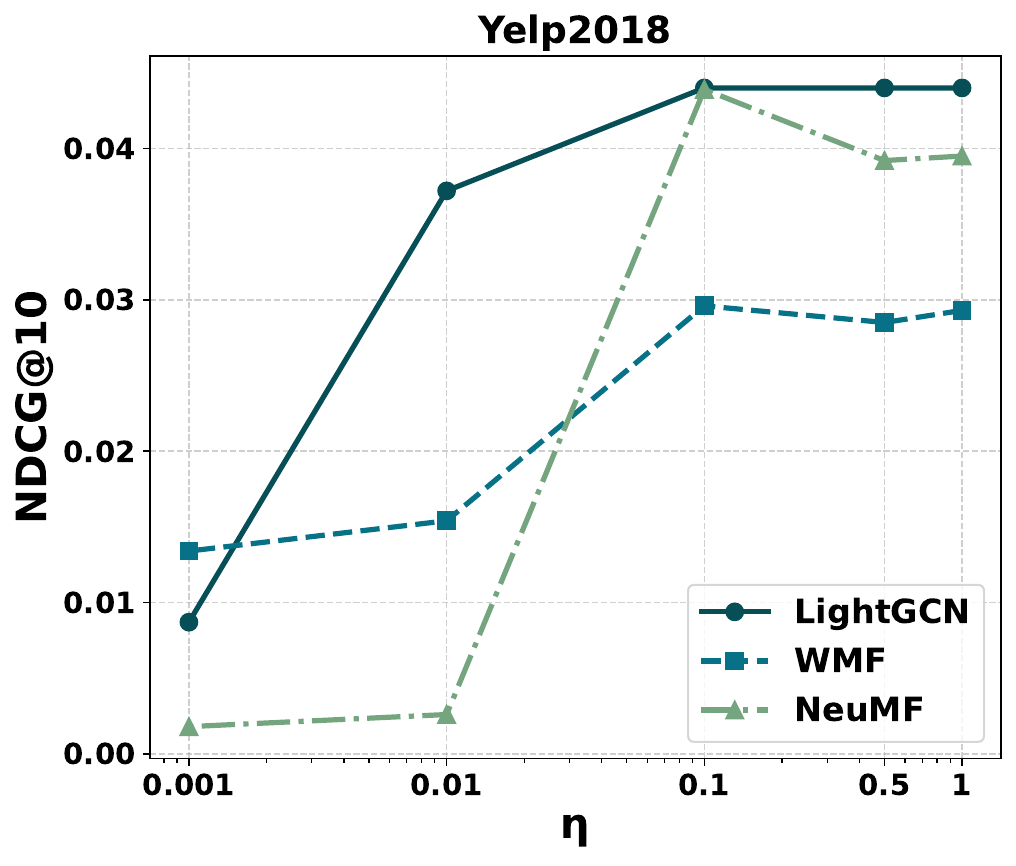}
        % \caption{Yelp2018}
        \label{fig:eta_ndcg_yelp2018}
    \end{subfigure}
    \begin{subfigure}[b]{0.33\columnwidth}
        \includegraphics[width=\columnwidth]{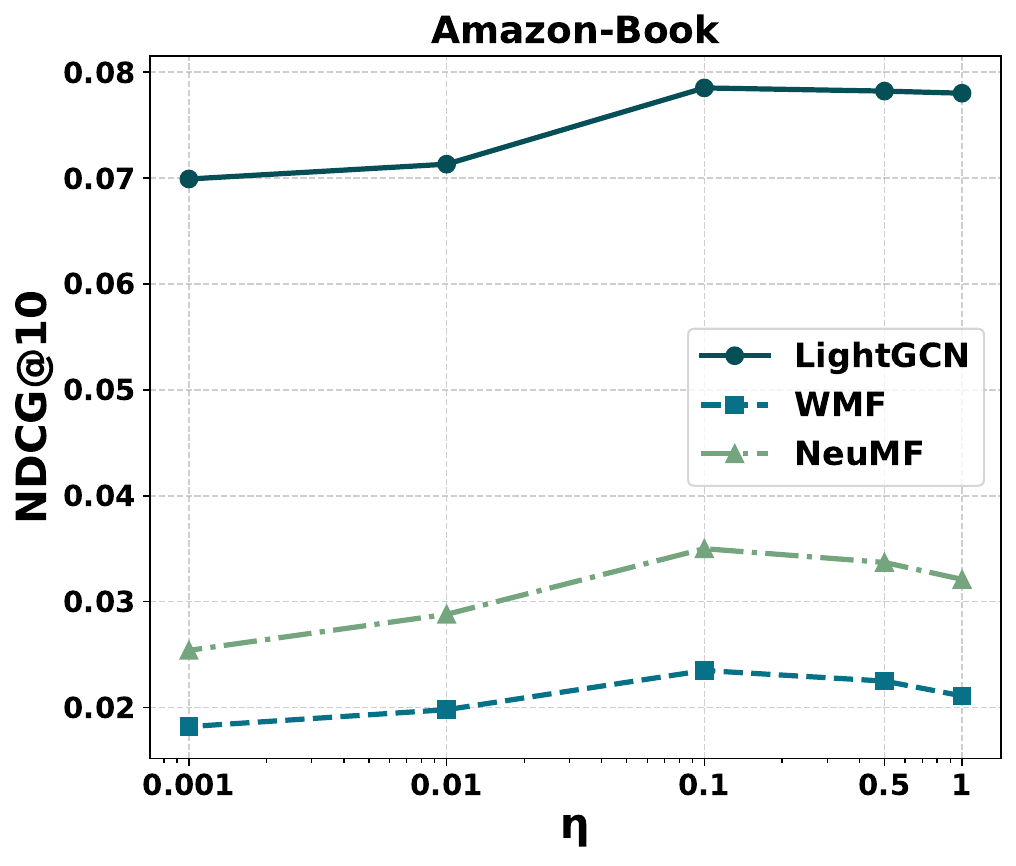}
        % \caption{Amazon-Book}
        \label{fig:eta_ndcg_amazon-book}
    \end{subfigure}
    \\
    \begin{subfigure}[b]{0.33\columnwidth}
        \includegraphics[width=\columnwidth]{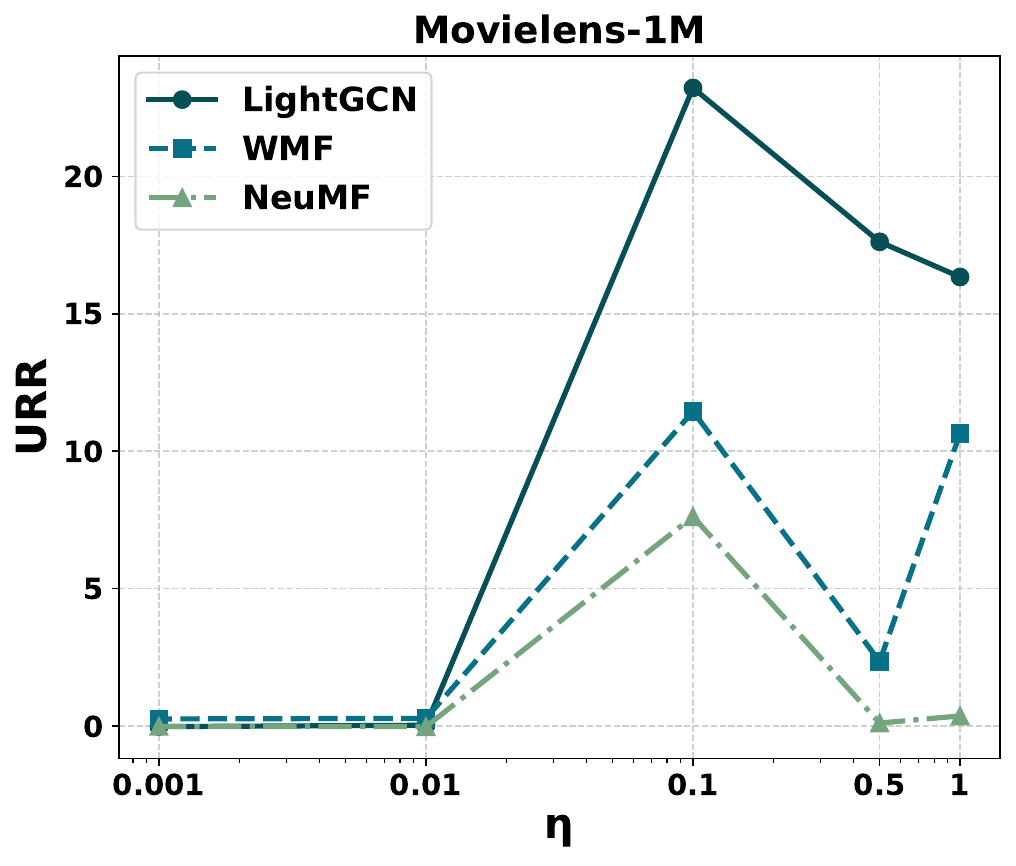}
        % \caption{MovieLens-1M}
        \label{fig:eta_urr_ml-1m}
    \end{subfigure}
    \begin{subfigure}[b]{0.33\columnwidth}
        \includegraphics[width=\columnwidth]{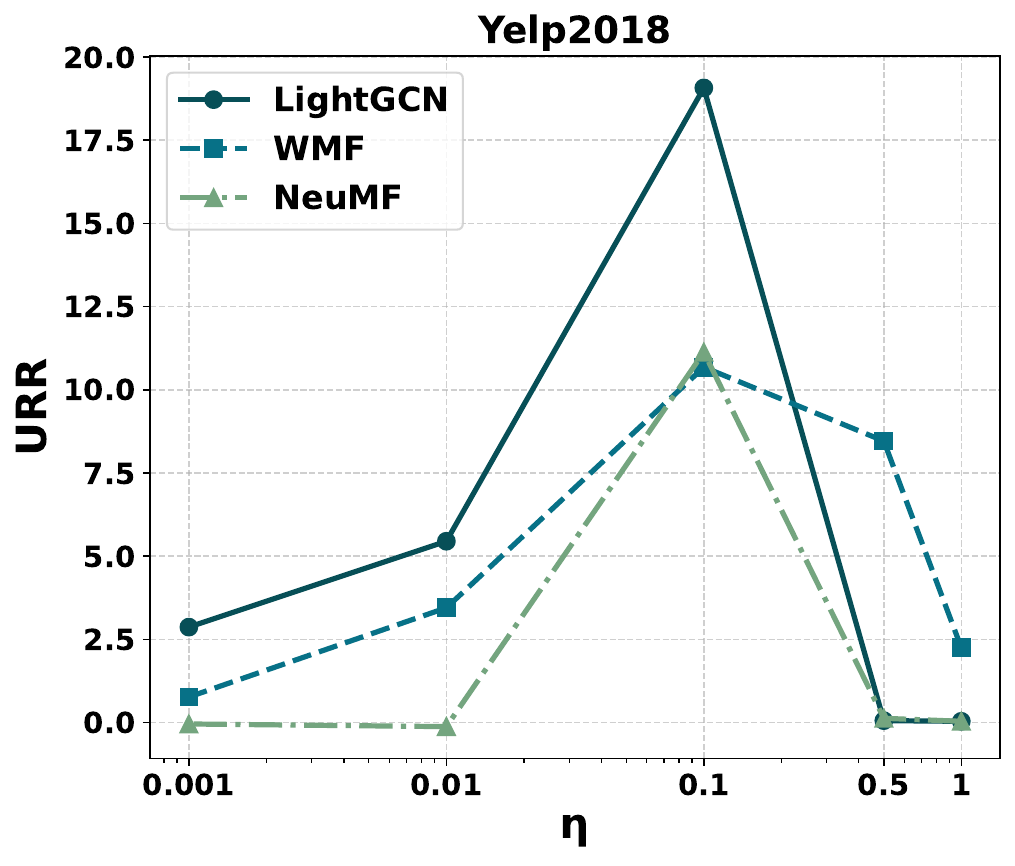}
        % \caption{Yelp2018}
        \label{fig:eta_urr_yelp2018}
    \end{subfigure}
    \begin{subfigure}[b]{0.33\columnwidth}
        \includegraphics[width=\columnwidth]{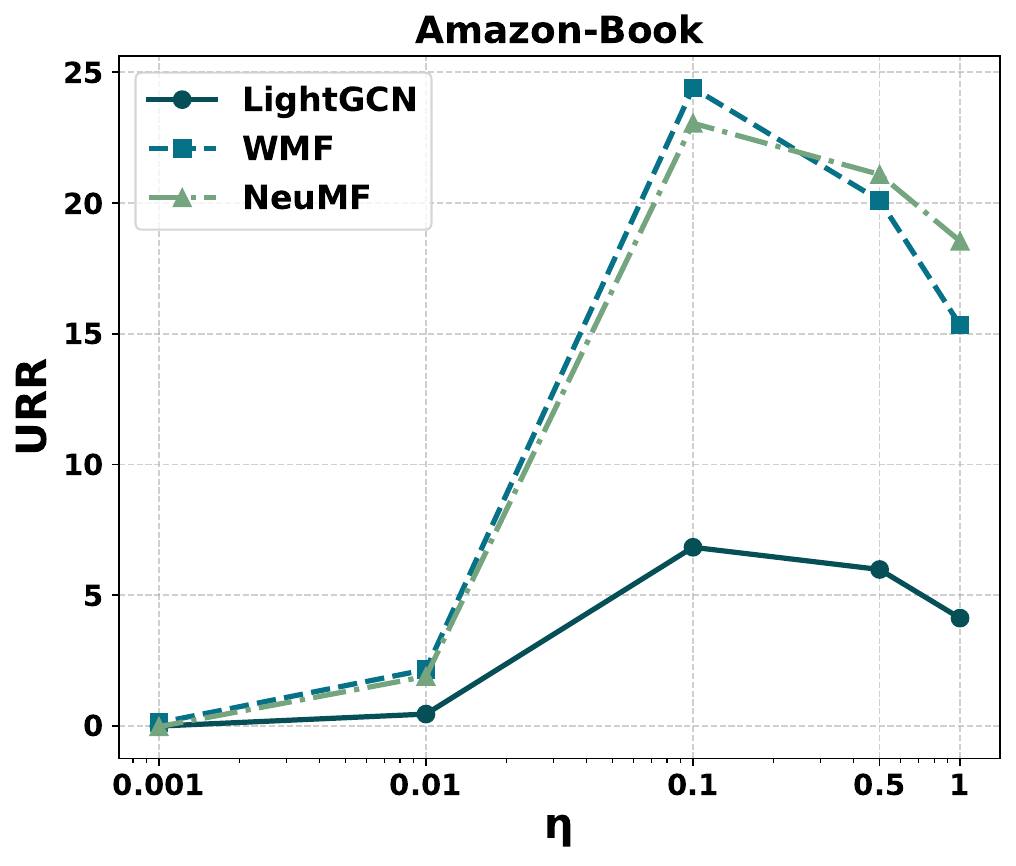}
        % \caption{Amazon-Book}
        \label{fig:eta_urr_amazon-book}
    \end{subfigure}

    \caption{Impact of scaling factor $\eta$ across three datasets and backbone models after randomly removing 5\% of items.}
    \label{fig:scaling_factor}
\end{figure}

The results in Figure~\ref{fig:scaling_factor} provide critical insights into how the scaling factor $\eta$ affects L2UnRank's performance across multiple dimensions:

\textbf{Optimal Range.} The value $\eta=0.1$ consistently delivers the best balance between recommendation quality and unlearning effectiveness across all datasets and backbone models. This setting achieves the highest URR values while maintaining competitive NDCG@10 scores, demonstrating that moderate scaling provides sufficient unlearning strength without excessive parameter perturbation.

\textbf{Ineffectiveness of Small Values.} Values of $\eta \leq 0.01$ result in negligible unlearning effectiveness, with URR values approaching zero or becoming negative. Such conservative parameter updates fail to meaningfully alter target item rankings, rendering the unlearning process ineffective.

\textbf{Instability of Large Values.} When $\eta \geq 0.5$, the method exhibits suboptimal or unstable behavior. Although these values occasionally achieve reasonable URR scores, they frequently cause substantial recommendation quality drops, particularly pronounced in MovieLens-1M and Yelp2018. Excessive parameter modifications disrupt learned representations and compromise the model's predictive accuracy on the retain set.

\textbf{Dataset-Specific Behavior.} Different datasets exhibit varying responsiveness to the scaling factor. MovieLens-1M demonstrates the most severe quality degradation under very small $\eta$ values, while Amazon-Book maintains relatively stable performance across the tested range. This variation suggests that optimal $\eta$ selection may depend on dataset characteristics, including interaction density and sparsity patterns.

\textbf{Model-Agnostic Consistency.} Despite architectural differences, all three backbone models (LightGCN, WMF, NeuMF) exhibit consistent trends regarding $\eta$ sensitivity, confirming the model-agnostic nature of our method. The uniform optimal performance at $\eta=0.1$ across different architectures validates our hyperparameter selection and demonstrates robust cross-model applicability.

\section{Comprehensive Experimental Analysis}
This section presents supplementary experimental results that provide deeper insights into L2UnRank's performance characteristics and robustness. We conduct comprehensive analyses across multiple dimensions, including scalability under varying unlearning ratios, complete performance comparisons across diverse backbone architectures, sensitivity analysis of key hyperparameters, and comparative evaluation against specialized graph-based unlearning methods. These additional experiments strengthen our understanding of L2UnRank's capabilities and validate its effectiveness across a broader range of scenarios than those presented in the main experimental section.

\subsection{Robustness Analysis}
To comprehensively evaluate the robustness of L2UnRank across varying unlearning scenarios and scales, we conduct extensive experiments under different unlearning ratios (2.5\%, 5\%, and 10\%) for both entity unlearning and interaction unlearning tasks. This analysis is crucial for understanding the method's stability and practical applicability in real-world scenarios where the volume of data to be unlearned may vary significantly.

As demonstrated in Table~\ref{tab:scalability_analysis_comprehensive}, L2UnRank exhibits remarkable robustness across different unlearning ratios and scenarios. The method consistently maintains high recommendation quality (NDCG@10) and superior unlearning effectiveness (URR) while achieving exceptional computational efficiency across all tested configurations. Notably, even when the unlearning ratio increases from 2.5\% to 10\%, L2UnRank sustains its performance advantages, demonstrating its stability in handling large-scale unlearning tasks without significant performance degradation.

\begin{table*}[ht]
\centering
\caption{Scalability and robustness comparison on the LightGCN backbone across different unlearning scenarios and ratios.}
\label{tab:scalability_analysis_comprehensive}
\resizebox{\textwidth}{!}{
\begin{tabular}{lll ccc ccc ccc}
\toprule
\multirow{2}{*}{\textbf{Scenario}} & \multirow{2}{*}{\textbf{Forget Ratio}} & \multirow{2}{*}{\textbf{Method}} & \multicolumn{3}{c}{\textbf{MovieLens-1M}} & \multicolumn{3}{c}{\textbf{Yelp2018}} & \multicolumn{3}{c}{\textbf{Amazon-Book}} \\
\cmidrule(lr){4-6} \cmidrule(lr){7-9} \cmidrule(lr){10-12}
& & & \textbf{NDCG@10} & \textbf{URR} & \textbf{Time(s)} & \textbf{NDCG@10} & \textbf{URR} & \textbf{Time(s)} & \textbf{NDCG@10} & \textbf{URR} & \textbf{Time(s)} \\
\midrule
\multirow{12}{*}{Entity Unlearning} & \multirow{4}{*}{2.50\%} 
& Retrain & 0.2041 & 13.82 & 48.95 & 0.0415 & 19.33 & 1380.16 & 0.0815 & 5.31 & 8810.34 \\
\cmidrule{3-12}
& & RecEraser & 0.2015 & 5.41 & 1255.3 & 0.0341 & 1.72 & 1510.8 & 0.0631 & 2.45 & 3988.1 \\
& & IFRU & 0.2031 & 1.35 & 48.66 & 0.0442 & 0.15 & 85.11 & 0.0791 & 0.85 & 185.33 \\
& & \textbf{Ours} & \textbf{0.2033} & \textbf{23.81} & \textbf{0.45} & \textbf{0.0446} & \textbf{19.95} & \textbf{0.48} & \textbf{0.0792} & \textbf{7.51} & \textbf{1.02} \\
\cmidrule(lr){2-12}
& \multirow{4}{*}{5\%}
& Retrain & 0.1975 & 13.61 & 47.18 & 0.0394 & 18.96 & 1342.52 & 0.0796 & 4.95 & 8600.60 \\
\cmidrule{3-12}
& & RecEraser & 0.1994 & 5.13 & 1476.6 & 0.0336 & 1.60 & 1757.0 & 0.0612 & 2.26 & 4250.9 \\
& & IFRU & 0.2020 & 1.22 & 60.57 & 0.0438 & 0.13 & 102.42 & 0.0783 & 0.79 & 199.47 \\
& & \textbf{Ours} & \textbf{0.2024} & \textbf{23.22} & \textbf{0.63} & \textbf{0.0440} & \textbf{19.07} & \textbf{0.65} & \textbf{0.0785} & \textbf{6.83} & \textbf{1.40} \\
\cmidrule(lr){2-12}
& \multirow{4}{*}{10\%}
& Retrain & 0.1885 & 14.95 & 45.30 & 0.0365 & 19.85 & 1295.11 & 0.0753 & 5.52 & 8350.15 \\
\cmidrule{3-12}
& & RecEraser & 0.1910 & 4.98 & 1860.1 & 0.0305 & 1.51 & 2133.5 & 0.0575 & 2.11 & 4890.7 \\
& & IFRU & \textbf{0.1975} & 1.19 & 85.12 & 0.0415 & 0.11 & 135.88 & 0.0760 & 0.75 & 230.19 \\
& & \textbf{Ours} & 0.1962 & \textbf{24.13} & \textbf{1.15} & \textbf{0.0417} & \textbf{20.24} & \textbf{1.21} & \textbf{0.0761} & \textbf{7.95} & \textbf{2.53} \\
\midrule
\multirow{12}{*}{Interaction Unlearning} & \multirow{4}{*}{2.50\%} 
& Retrain & 0.2066 & 12.51 & 49.53 & 0.0461 & 16.55 & 1392.14 & 0.0821 & 4.88 & 8850.22 \\
\cmidrule{3-12}
& & RecEraser & 0.2028 & 4.88 & 1105.7 & 0.0355 & 1.51 & 1430.2 & 0.0645 & 2.13 & 3870.4 \\
& & IFRU & 0.2045 & 1.15 & 40.21 & \textbf{0.0453} & 0.14 & 75.34 & \textbf{0.0805} & 0.72 & 175.83 \\
& & \textbf{Ours} & \textbf{0.2053} & \textbf{21.89} & \textbf{0.31} & 0.0451 & \textbf{18.15} & \textbf{0.35} & 0.0801 & \textbf{6.65} & \textbf{0.88} \\
\cmidrule(lr){2-12}
& \multirow{4}{*}{5\%}
& Retrain & 0.2050 & 12.33 & 48.19 & 0.0455 & 16.31 & 1366.88 & 0.0813 & 4.51 & 8699.12 \\
\cmidrule{3-12}
& & RecEraser & 0.2001 & 4.71 & 1321.4 & 0.0342 & 1.44 & 1698.2 & 0.0621 & 2.01 & 4130.5 \\
& & IFRU & 0.2038 & 1.08 & 55.73 & 0.0448 & 0.12 & 98.72 & \textbf{0.0799} & 0.68 & 191.40 \\
& & \textbf{Ours} & \textbf{0.2051} & \textbf{21.06} & \textbf{0.59} & \textbf{0.0449} & \textbf{17.63} & \textbf{0.62} & 0.0798 & \textbf{6.17} & \textbf{1.35} \\
\cmidrule(lr){2-12}
& \multirow{4}{*}{10\%}
& Retrain & 0.1995 & 13.15 & 46.88 & 0.0428 & 17.11 & 1310.50 & 0.0780 & 4.85 & 8412.30 \\
\cmidrule{3-12}
& & RecEraser & 0.1945 & 4.55 & 1715.2 & 0.0317 & 1.38 & 2005.1 & 0.0585 & 1.95 & 4755.1 \\
& & IFRU & 0.1990 & 1.05 & 78.34 & 0.0422 & 0.10 & 125.19 & 0.0772 & 0.63 & 225.88 \\
& & \textbf{Ours} & \textbf{0.2008} & \textbf{22.05} & \textbf{1.09} & \textbf{0.0426} & \textbf{18.92} & \textbf{1.18} & \textbf{0.0775} & \textbf{6.98} & \textbf{2.41} \\
\bottomrule
\end{tabular}
}
\end{table*}

\subsection{Comprehensive Unlearning Effectiveness Analysis}
To provide a complete assessment of L2UnRank's performance, we present comprehensive experimental results that extend beyond the main paper's analysis. Table~\ref{tab:unlearning_performance} includes additional backbone models (WMF and NeuMF) and datasets (Amazon-Book with LightGCN) that were not fully presented in the main experimental section due to space constraints.

The comprehensive results consistently validate our method's superiority across diverse recommendation architectures. For LightGCN, L2UnRank achieves URR values of 23.22, 19.07, and 6.83 on MovieLens-1M, Yelp2018, and Amazon-Book respectively, substantially outperforming all baseline methods. Particularly noteworthy is the comparison with IFRU, which demonstrates significantly lower URR values (1.22, 0.13, and 0.79 respectively), indicating L2UnRank's superior ability to effectively remove the influence of forgotten entities from model predictions.

The results also demonstrate L2UnRank's model-agnostic nature. For WMF, our method achieves competitive URR values of 11.45, 10.68, and 24.40 across the three datasets, while maintaining acceptable FPR levels. For NeuMF, L2UnRank excels particularly on the Amazon-Book dataset with a URR of 23.05, highlighting its effectiveness across different neural architectures. These comprehensive results reinforce the conclusions presented in Section~\ref{sec_performance_comparison}, confirming L2UnRank's superior performance in terms of both unlearning effectiveness and computational efficiency across diverse recommendation scenarios.

\begin{table*}[t]
    \centering
    \footnotesize
    \caption{Complete unlearning effectiveness comparison after randomly deleting 5\% of items.}
    \label{tab:unlearning_performance}
    \begin{tabular}{@{}llrcrcrc@{}}
        \toprule
        \multirow{2}{*}{\textbf{Backbone}} & \multirow{2}{*}{\textbf{Method}} & \multicolumn{2}{c}{\textbf{MovieLens-1M}} & \multicolumn{2}{c}{\textbf{Yelp2018}} & \multicolumn{2}{c}{\textbf{Amazon-Book}} \\ 
        \cmidrule(lr){3-4} \cmidrule(lr){5-6} \cmidrule(lr){7-8}
        & & URR & FPR & URR & FPR & URR & FPR \\ 
        \midrule
        \multirow{6}{*}{\textbf{LightGCN}} & Retrain &13.61 &0.0574 &18.96 &0.0137 &4.95 &0.0883 \\
        & CertifiedRemoval &5.39 &0.0825 &15.32 &\textbf{0.0771} &2.96 &0.0564 \\
        \cmidrule{2-8}
        & SISA &5.13 &\textbf{0.0471} &1.60 &\textbf{0.0313} &2.26 &0.0201 \\
        & RecEraser &5.37 &0.0261 &14.00 &0.0107 &3.94 &0.0530 \\
        & IFRU &1.22 &0.0210 &0.13 &0.0113 &0.79 &0.0583 \\
        & Ours &\textbf{23.22} &0.0458 &\textbf{19.07} &0.0143 &\textbf{6.83} &\textbf{0.0661} \\ 
        \midrule
        \multirow{6}{*}{\textbf{WMF}} & Retrain &8.08 &0.0765 &10.66 &0.2584 &26.95 &0.2060 \\
        & CertifiedRemoval &5.83 &0.0868 &13.59 &0.1888 &21.91 &0.6041 \\
        \cmidrule{2-8}
        & SISA &5.61 &0.0740 &13.55 &0.2540 &21.77 &0.3062 \\
        & RecEraser &5.78 &\textbf{0.0987} &13.65 &0.2041 &21.84 &0.2306 \\
        & IFRU &0.03 &0.0355 &1.54 &0.0721 &5.21 &0.1344 \\
        & Ours &\textbf{11.45} &0.0402 &\textbf{15.68} &\textbf{0.2226} &\textbf{24.40} &\textbf{0.4800} \\ 
        \midrule
        \multirow{6}{*}{\textbf{NeuMF}} & Retrain &0.58 &0.0857 &11.05 &0.3075 &3.13 &0.2742 \\
        & CertifiedRemoval &0.04 &0.1847 &8.32 &0.3737 &1.32 &0.3021 \\
        \cmidrule{2-8}
        & SISA &1.65 &0.1115 &6.89 &0.0721 &0.59 &0.0569 \\
        & RecEraser &4.52 &0.0634 &9.17 &0.0305 &3.30 &\textbf{0.1805} \\
        & IFRU &0.03 &0.0240 &0.66 &0.0132 &0.69 &0.1422 \\
        & Ours &\textbf{7.62} &\textbf{0.1620} &\textbf{14.13} &\textbf{0.0343} &\textbf{23.05} &0.1353 \\ 
        \bottomrule
    \end{tabular}
\end{table*}

\subsection{Impact of Influenced Scope Size $p$}

% \begin{figure}[t]
%     \centering
%     \includegraphics[width=0.4\linewidth]{figs/p/p_amazon-book.pdf}
%     \caption{Effect of influenced scope size parameter $p$ on unlearning performance and recommendation accuracy after randomly deleting 5\% of items.}
%     \label{fig:p_amazon-book}
% \end{figure}

\begin{table}[htbp!]
\centering
\footnotesize
\caption{Impact of scope size $p$ on performance (NDCG@10, URR) and efficiency (Time in seconds) for Amazon-Book dataset across different backbone models.}
\label{tab:p_amazon-book}
\begin{tabular}{l ccc ccc ccc}
\toprule
\multirow{2}{*}{\textbf{p}} & \multicolumn{3}{c}{\textbf{LightGCN}} & \multicolumn{3}{c}{\textbf{WMF}} & \multicolumn{3}{c}{\textbf{NeuMF}} \\
\cmidrule(lr){2-4} \cmidrule(lr){5-7} \cmidrule(lr){8-10}
& \textbf{NDCG@10} & \textbf{URR} & \textbf{Time(s)} & \textbf{NDCG@10} & \textbf{URR} & \textbf{Time(s)} & \textbf{NDCG@10} & \textbf{URR} & \textbf{Time(s)} \\
\midrule
0 & 0.0671 & 4.55  & \textbf{1.30} & \textbf{0.0235} & \textbf{33.10} & \textbf{0.93} & \textbf{0.0350} & 23.05 & \textbf{1.73} \\
1 & \textbf{0.0805} & \textbf{6.83}  & 1.40 & 0.0204 & 24.40 & 0.98 & 0.0321 & \textbf{24.55} & 1.85 \\
2 & 0.0803 & 5.88  & 1.58 & 0.0205 & 12.30 & 1.01 & 0.0288 & 19.82 & 1.96 \\
3 & 0.0785 & 5.79  & 1.88 & 0.0147 & 12.60 & 1.54 & 0.0254 & 17.65 & 2.40 \\
\bottomrule
\end{tabular}
\end{table}

In Section~\ref{sec_hyperparameter_analysis}, we analyzed the effect of the influenced scope size parameter $p$ on the performance of L2UnRank using the MovieLens-1M and Yelp2018 datasets, as shown in Figure~\ref{fig:influence_scope}. To further validate the robustness of this parameter, we conducted the same experiment on the Amazon-Book dataset. The results, illustrated in Table~\ref{tab:p_amazon-book}, reveal a consistent trend across all datasets.

For LightGCN, optimal recommendation accuracy and unlearning effectiveness are consistently achieved at $p=1$. This aligns with our expectation that for a graph-based model like LightGCN, which relies on message passing to capture the collaborative filtering signal, the 1-hop neighborhood is sufficient to encapsulate the most direct and critical collaborative effects. A setting of $p=0$ is overly restrictive as it fails to account for neighbor influences, whereas expanding the scope to $p>1$ introduces computational overhead and noise from more distant, potentially irrelevant nodes, leading to diminishing returns or even performance degradation.

Conversely, for WMF and NeuMF, optimal performance is generally observed at $p=0$. These models rely on interactions between latent factors of users and items rather than explicit graph propagation. Consequently, their predictions are primarily influenced by direct interactions. Expanding the influenced scope beyond the directly affected entities ($p>0$) not only incurs unnecessary computational cost but also introduces entities with weaker relevance to the unlearning task, thereby interfering with the precision of the parameter update and degrading both recommendation utility and unlearning effectiveness.

These supplementary results reinforce our conclusion that the optimal influenced scope $p$ is intrinsically dependent on the backbone model's architecture. The ability of our method to achieve optimal performance with small $p$ values ($p=1$ for graph-based models and $p=0$ for latent factor models) underscores the precision and efficiency of L2UnRank in identifying the most critical influenced scope.

\subsection{Impact of Influence Balancing Parameter $\alpha$}

\begin{figure}[t]
    \centering
    \begin{subfigure}[b]{0.33\columnwidth}
        \includegraphics[width=\columnwidth]{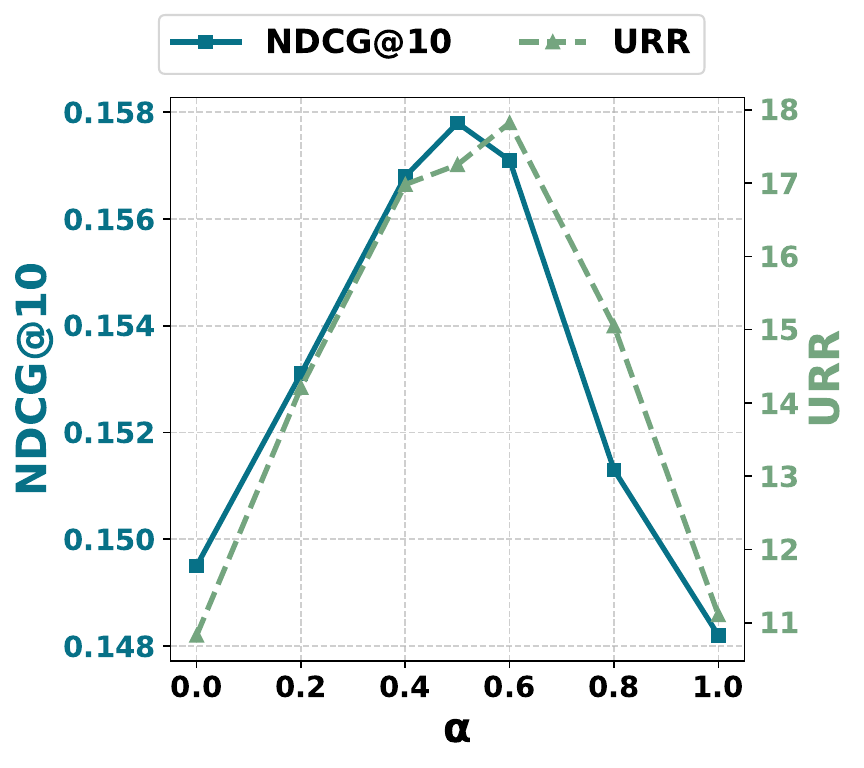}
        % \caption{MovieLens-1M}
        \label{fig:alpha_wmf_ml-1m}
    \end{subfigure}
    \hfill
    \begin{subfigure}[b]{0.33\columnwidth}
        \includegraphics[width=\columnwidth]{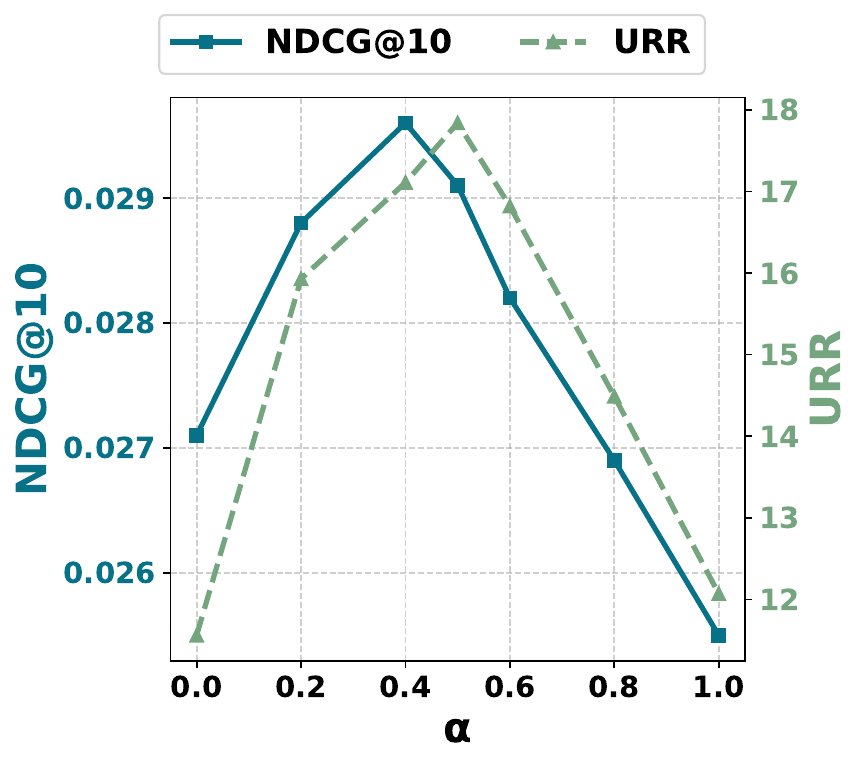}
        % \caption{Yelp2018}
        \label{fig:alpha_wmf_yelp2018}
    \end{subfigure}
    \begin{subfigure}[b]{0.33\columnwidth}
        \includegraphics[width=\columnwidth]{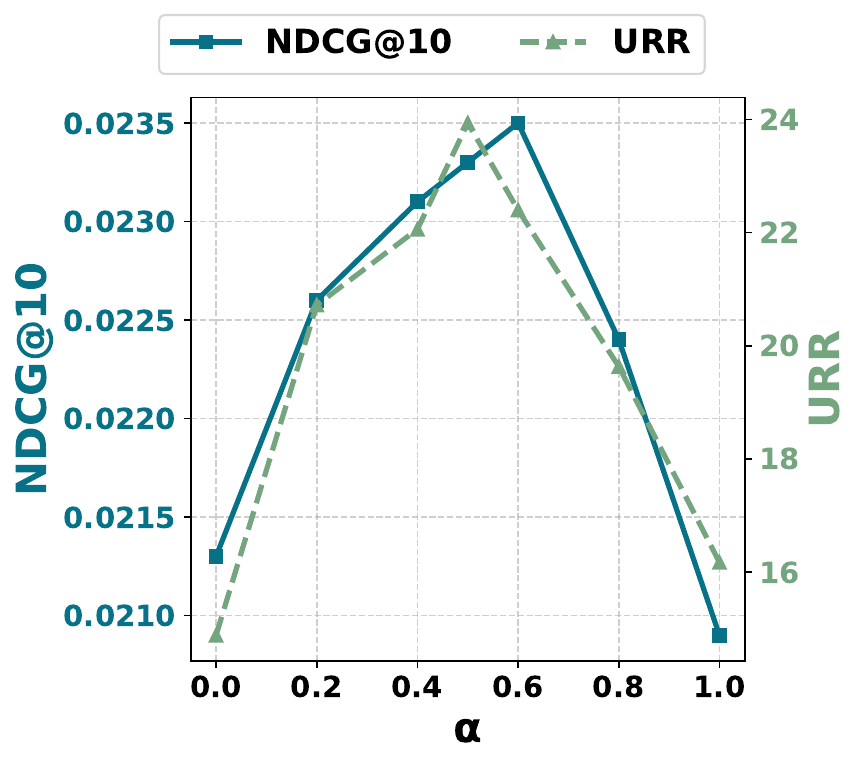}
        % \caption{Amazon-Book}
        \label{fig:alpha_wmf_amazon-book}
    \end{subfigure}
    \caption{Impact of influence balancing parameter $\alpha$ using \textbf{WMF} across three datasets after randomly removing 5\% of items.}
    \label{fig:alpha_wmf}
\end{figure}

\begin{figure}[t]
    \centering
    \begin{subfigure}[b]{0.33\columnwidth}
        \includegraphics[width=\columnwidth]{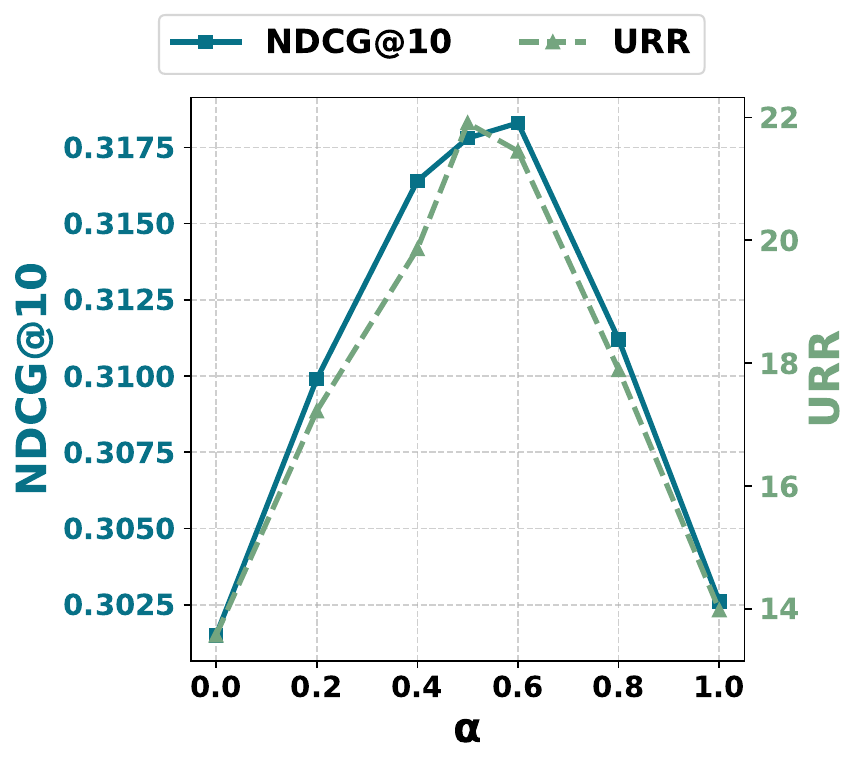}
        % \caption{MovieLens-1M}
        \label{fig:alpha_neumf_ml-1m}
    \end{subfigure}
    \hfill
    \begin{subfigure}[b]{0.33\columnwidth}
        \includegraphics[width=\columnwidth]{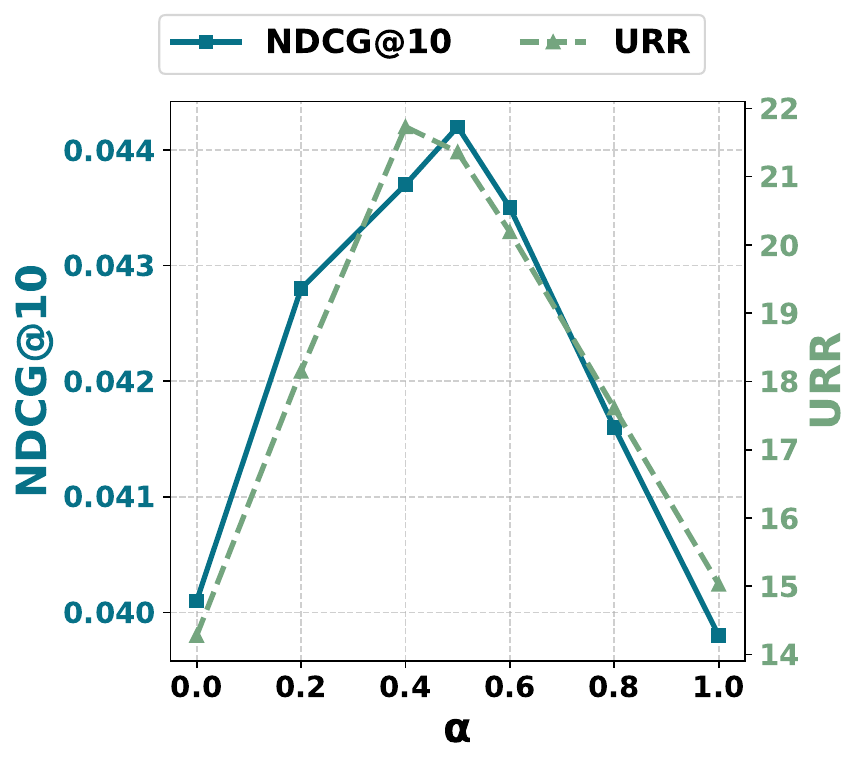}
        % \caption{Yelp2018}
        \label{fig:alpha_neumf_yelp2018}
    \end{subfigure}
    \begin{subfigure}[b]{0.33\columnwidth}
        \includegraphics[width=\columnwidth]{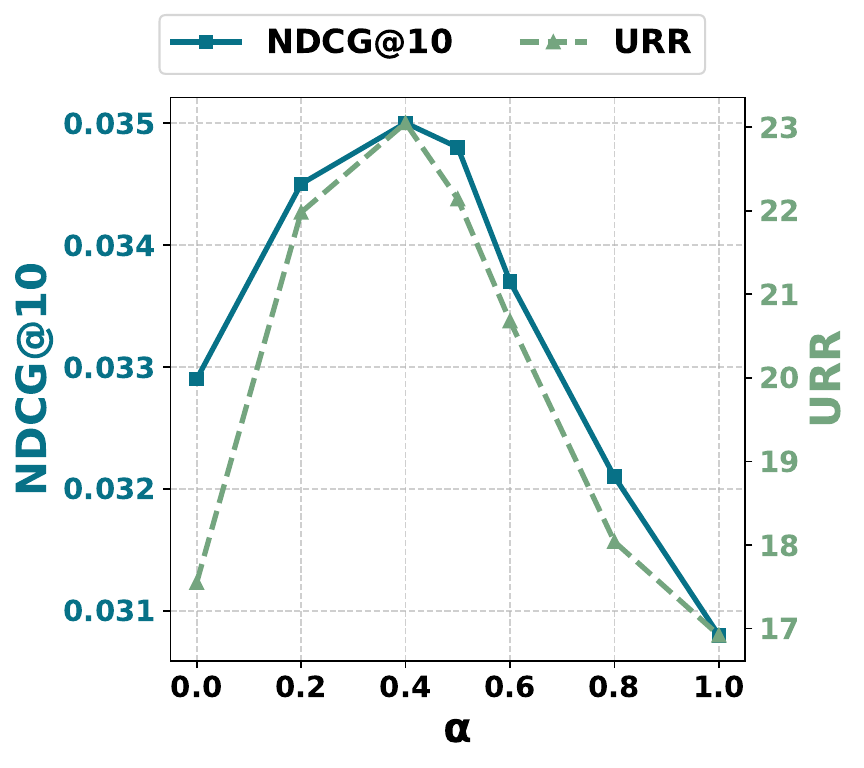}
        % \caption{Amazon-Book}
        \label{fig:alpha_neumf_amazon-book}
    \end{subfigure}
    \caption{Impact of influence balancing parameter $\alpha$ using \textbf{NeuMF} across three datasets after randomly removing 5\% of items.}
    \label{fig:alpha_neumf}
\end{figure}

In Section~\ref{sec_hyperparameter_analysis}, we demonstrated the importance of balancing structural and semantic influence through analysis of the parameter $\alpha$ for the LightGCN model. To establish the generalizability of this design principle across different recommendation architectures, we extend this sensitivity analysis to WMF and NeuMF models. The results, presented in Figures~\ref{fig:alpha_wmf} and~\ref{fig:alpha_neumf}, provide compelling evidence for the universal applicability of our influence balancing strategy.

The experimental findings reveal that the complementary nature of structural and semantic influence constitutes a model-agnostic principle fundamental to effective unlearning. Across all tested architectures (WMF and NeuMF), consistent with our observations for LightGCN, performance degrades substantially at extreme values of $\alpha$ (approaching 0 or 1) and achieves optimal results within moderate ranges, typically between 0.4 and 0.6. This pattern demonstrates that relying on a single source of influence—whether purely semantic ($\alpha=0$) or purely structural ($\alpha=1$)—is fundamentally insufficient for effective unlearning.

When $\alpha$ approaches 0, the model depends exclusively on semantic similarity, disregarding the structural importance of entities within the user-item interaction graph. This limitation becomes particularly problematic in sparse datasets where semantic signals may be insufficient to accurately identify all relevant entities for unlearning. Conversely, when $\alpha$ approaches 1, the model relies solely on structural connectivity patterns, ignoring content-level relevance. This approach may inadvertently amplify the influence of popular but contextually inappropriate items, leading to suboptimal unlearning outcomes.

The consistent optimal performance achieved with balanced $\alpha$ values across different model architectures validates our hypothesis that L2UnRank's effectiveness stems from creating a comprehensive and robust entity influence distribution by synergistically integrating both structural and semantic perspectives. For instance, on the Yelp2018 dataset, WMF achieves peak URR at $\alpha=0.5$, while NeuMF obtains optimal URR and NDCG@10 at $\alpha=0.4$ and $\alpha=0.5$, respectively. This ability to achieve an effective balance between preserving model utility and ensuring unlearning effectiveness across diverse architectures confirms the fundamental importance and universal applicability of our Fine-Grained Influence Quantification module.

These cross-architecture results provide strong empirical evidence that our influence balancing strategy is not merely an architectural-specific optimization but represents a fundamental principle for effective recommendation unlearning. The consistent performance patterns across WMF, NeuMF, and LightGCN underscore that this component constitutes an essential and universally applicable element of the L2UnRank framework.

\section{Comparison with Graph-based Unlearning Methods}
To provide a comprehensive evaluation of L2UnRank's effectiveness, we compare our method with GSGCF-RU (General Strategy Graph Collaborative Filtering for Recommendation Unlearning)~\cite{hao2024general}, a state-of-the-art graph-based unlearning approach specifically designed for collaborative filtering recommendation systems.

\subsection{GSGCF-RU Overview}
GSGCF-RU represents a learning-based recommendation unlearning method that leverages two core principles: Unlearning Edge Consistency (UEC) and Feature Representation Consistency (FRC). The method employs a Learnable Deletion Operator (LDO) with local adjustment strategies to achieve unlearning in Graph Neural Network (GNN) models. This approach is particularly relevant for comparison as it targets the same domain of graph-based collaborative filtering systems.

\subsection{Experimental Setup}
Following the original paper's experimental protocol, we configure GSGCF-RU with 4 LDO layers, employing the Adam optimizer with a learning rate of 0.01, training for 100 epochs with a residual coefficient of 0.1. The hyperparameter $\lambda$ for balancing UEC and FRC losses is set to 0.5~\cite{cheng2023gnndelete}, and the L2 regularization coefficient is set to $1 \times 10^{-5}$. Since GSGCF-RU is specifically designed for LightGCN, we conduct a fair comparison using identical LightGCN model configurations for both methods.

\subsection{Comparative Results and Analysis}

\begin{table}[htbp]
  \centering
  \footnotesize
  \caption{Comparison between L2UnRank and GSGCF-RU.}
  \label{tab:gsgcfru_comparison}
  \begin{tabular}{llcccc}
    \toprule
    Dataset & Method & NDCG@10 & Time & URR & FPR \\
    \midrule
    \multirow{3}{*}{MovieLens-1M} & Retrain & 0.1975 & 47.18 & 13.61 & \textbf{0.0574} \\
                                  & GSGCF-RU & 0.1917 & 56.63 & 0.82 & 0.0473 \\
                                  & Ours & \textbf{0.2024} & \textbf{0.63} & \textbf{23.22} & 0.0458 \\
    \midrule
    \multirow{3}{*}{Yelp2018} & Retrain & 0.0394 & 1342.52 & 18.96 & 0.0137 \\
                              & GSGCF-RU & 0.0439 & 89.87 & 0.38 & 0.0113 \\
                              & Ours & \textbf{0.0440} & \textbf{0.65} & \textbf{19.07} & \textbf{0.0143} \\
    \midrule
    \multirow{3}{*}{Amazon-Book} & Retrain & \textbf{0.0796} & 8600.60 & 4.95 & \textbf{0.0883} \\
                                 & GSGCF-RU & 0.0736 & 164.85 & 0.79 & 0.0687 \\
                                 & Ours & 0.0785 & \textbf{1.40} & \textbf{6.83} & 0.0661 \\
    \bottomrule
  \end{tabular}
\end{table}

The experimental results presented in Table~\ref{tab:gsgcfru_comparison} reveal significant performance disparities between L2UnRank and GSGCF-RU across all evaluation metrics and datasets. L2UnRank demonstrates substantial superiority in both unlearning effectiveness and computational efficiency.

In terms of unlearning effectiveness, measured by the Unlearning Removal Rate (URR), L2UnRank achieves remarkable improvements over GSGCF-RU. On the MovieLens-1M dataset, L2UnRank attains a URR of 23.22 compared to GSGCF-RU's 0.82, representing a 28-fold improvement in the method's ability to degrade target item rankings. Similar patterns are observed across other datasets: on Yelp2018, L2UnRank achieves 19.07 URR versus GSGCF-RU's 0.38, and on Amazon-Book, the corresponding values are 6.83 and 0.79 respectively.

Regarding recommendation quality preservation, L2UnRank maintains competitive or superior NDCG@10 scores. On MovieLens-1M, L2UnRank achieves 0.2024 compared to GSGCF-RU's 0.1917, indicating better utility preservation. This demonstrates L2UnRank's ability to achieve effective unlearning without compromising recommendation accuracy.

The computational efficiency comparison reveals L2UnRank's substantial advantage in practical applicability. L2UnRank requires only 0.63 seconds on MovieLens-1M compared to GSGCF-RU's 56.63 seconds, representing approximately a 90-fold speedup. This efficiency gain is consistent across all datasets, with L2UnRank requiring 0.65 seconds versus 89.87 seconds on Yelp2018, and 1.40 seconds versus 164.85 seconds on Amazon-Book.

These comprehensive results validate that L2UnRank, despite being a model-agnostic approach, maintains exceptional effectiveness and efficiency in graph-based collaborative filtering recommendation unlearning tasks. The superior performance across multiple dimensions—unranking effectiveness, utility preservation, and computational efficiency—demonstrates the robustness and practical value of our proposed method.

\section{Conjugate Gradient Convergence and Stability}
\label{app_cg}

The Conjugate Gradient (CG) algorithm is central to our L2UnRank framework for efficiently solving the linear system $H\Delta\Theta = -\mathbf{g}$ without explicitly computing the Hessian inverse. Understanding its convergence properties and numerical stability is crucial for the reliable application of our method in practice.

\subsection{Algorithm Foundation}

The CG method iteratively solves the linear system $Ax = b$ where $A$ is a symmetric positive definite matrix. In our context, $A$ corresponds to the Hessian matrix $H$, $x$ represents the parameter update $\Delta\Theta$, and $b$ is the negative gradient $-\mathbf{g}$. The algorithm maintains conjugate search directions and performs exact line searches along these directions.

The basic CG iteration can be summarized as:
\begin{align}
\mathbf{r}_k &= \mathbf{r}_{k-1} - \alpha_k A\mathbf{p}_k \\
\mathbf{p}_{k+1} &= \mathbf{r}_k + \beta_k \mathbf{p}_k \\
\alpha_k &= \frac{\mathbf{r}_{k-1}^T \mathbf{r}_{k-1}}{\mathbf{p}_k^T A \mathbf{p}_k}, \quad \beta_k = \frac{\mathbf{r}_k^T \mathbf{r}_k}{\mathbf{r}_{k-1}^T \mathbf{r}_{k-1}}
\end{align}
where $\mathbf{r}_k$ is the residual, $\mathbf{p}_k$ is the search direction, and $\alpha_k$, $\beta_k$ are step size parameters.

\subsection{Theoretical Convergence Analysis}

\textbf{Finite Convergence Property.} For an $n$-dimensional quadratic function, CG theoretically converges to the exact solution within at most $n$ iterations under exact arithmetic. This finite convergence property makes CG particularly attractive for solving linear systems arising from quadratic optimization problems.

\textbf{Convergence Rate.} In practice, the convergence rate is closely related to the condition number $\kappa(A) = \lambda_{\max}/\lambda_{\min}$ of matrix $A$, where $\lambda_{\max}$ and $\lambda_{\min}$ are the largest and smallest eigenvalues, respectively. The error reduction follows:
\begin{equation}
\frac{\|\mathbf{x}_k - \mathbf{x}^*\|_A}{\|\mathbf{x}_0 - \mathbf{x}^*\|_A} \leq 2\left(\frac{\sqrt{\kappa(A)} - 1}{\sqrt{\kappa(A)} + 1}\right)^k
\end{equation}
where $\|\cdot\|_A$ denotes the $A$-norm and $\mathbf{x}^*$ is the exact solution.

This bound reveals that CG exhibits superlinear convergence when $\kappa(A)$ is moderate, but may converge slowly for ill-conditioned systems with large condition numbers.

\textbf{Clustering Effect.} When eigenvalues of $A$ are clustered, CG often converges much faster than the worst-case bound suggests. This clustering effect is particularly beneficial in machine learning applications where the Hessian often has favorable spectral properties.

\subsection{Numerical Stability Considerations}

\textbf{Loss of Orthogonality.} In finite-precision arithmetic, the theoretical conjugacy of search directions gradually deteriorates due to rounding errors. The loss of orthogonality among residual vectors can lead to convergence stagnation and reduced numerical accuracy.

\textbf{Conditioning Impact.} Ill-conditioned matrices amplify the effects of rounding errors, leading to poor convergence behavior and numerical instability. The condition number directly affects both convergence speed and stability, making preconditioning essential for challenging problems.

\textbf{Restart Strategy.} To mitigate numerical instability, periodic restarts can be employed. Restarting the algorithm every $n$ iterations (or when certain numerical criteria are violated) helps restore the conjugacy properties and improve overall stability.

\subsection{Practical Implementation in L2UnRank}

In our L2UnRank framework, several factors contribute to the effectiveness and stability of the CG implementation:

\textbf{Localized Scope.} By restricting updates to the influenced scope $D_{inf}$, we effectively work with a subset of parameters, reducing the dimensionality of the linear system and improving conditioning properties.

\textbf{Hessian-Vector Products.} We compute $H\mathbf{v}$ products using automatic differentiation without explicitly forming the Hessian matrix. This approach reduces memory requirements and computational overhead while maintaining numerical accuracy.

\textbf{Early Termination.} We employ adaptive termination criteria based on relative residual reduction:
\begin{equation}
\frac{\|\mathbf{r}_k\|}{\|\mathbf{r}_0\|} < \epsilon_{tol}
\end{equation}
where $\epsilon_{tol}$ is a user-specified tolerance (typically $10^{-6}$ to $10^{-8}$).

\textbf{Preconditioning Considerations.} While our current implementation uses standard CG, the framework can be extended to incorporate preconditioning strategies such as incomplete Cholesky decomposition or diagonal preconditioning to further improve convergence for challenging problems.

\subsection{Convergence Behavior Analysis}

\begin{figure}[htbp]
    \centering
    \includegraphics[width=\columnwidth]{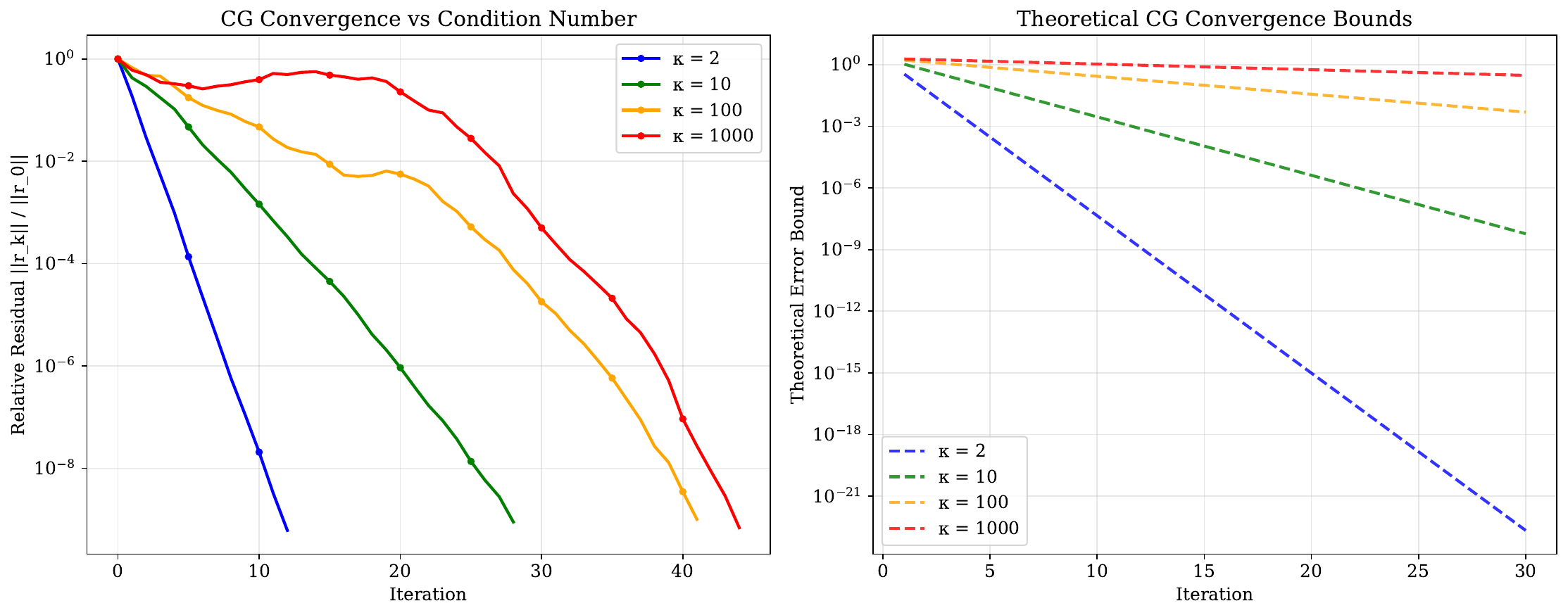}
    \caption{Convergence behavior of CG algorithm for different condition numbers. Left: Actual residual reduction showing superlinear convergence. Right: Theoretical convergence bounds demonstrating the effect of condition number on convergence rate.}
    \label{fig:cg_convergence}
\end{figure}

Figure~\ref{fig:cg_convergence} illustrates the typical convergence behavior of CG in our framework across different scenarios. The convergence curves demonstrate several key characteristics:

\textbf{Superlinear Convergence.} Most cases exhibit the characteristic superlinear convergence pattern: initially slow progress followed by rapid convergence in later iterations, as shown in the left panel of Figure~\ref{fig:cg_convergence}.

\textbf{Condition Number Effects.} Well-conditioned problems (small influenced scopes, regular data distributions) converge rapidly, while ill-conditioned cases require more iterations but still achieve acceptable solutions within reasonable computational budgets. The theoretical convergence bounds, illustrated in the right panel, confirm this relationship.

\textbf{Problem-Specific Behavior.} Different datasets and model architectures exhibit varying convergence characteristics, reflecting the underlying problem structure and data properties.

\begin{figure}[htbp]
    \centering
    \includegraphics[width=\columnwidth]{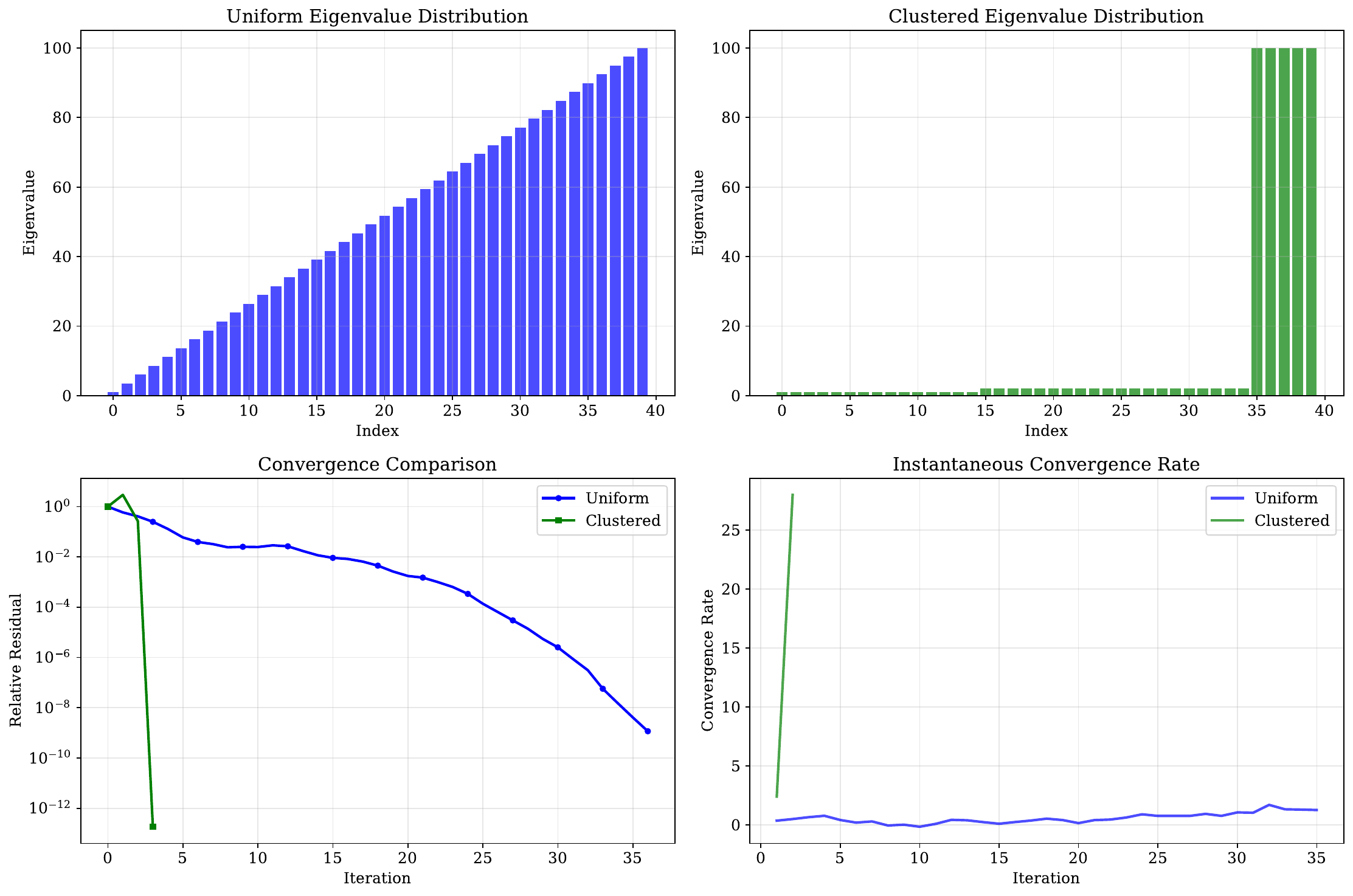}
    \caption{Effect of eigenvalue clustering on CG convergence. Top row shows eigenvalue distributions (uniform vs. clustered). Bottom row compares convergence behavior and instantaneous convergence rates, demonstrating faster convergence for clustered eigenvalues.}
    \label{fig:cg_eigenvalue_clustering}
\end{figure}

Figure~\ref{fig:cg_eigenvalue_clustering} demonstrates the beneficial effect of eigenvalue clustering on CG convergence. When eigenvalues are clustered (as often occurs in recommendation systems due to the low-rank nature of user-item interaction matrices), CG converges significantly faster than the worst-case theoretical bound suggests.

\begin{figure}[htbp]
    \centering
    \includegraphics[width=\columnwidth]{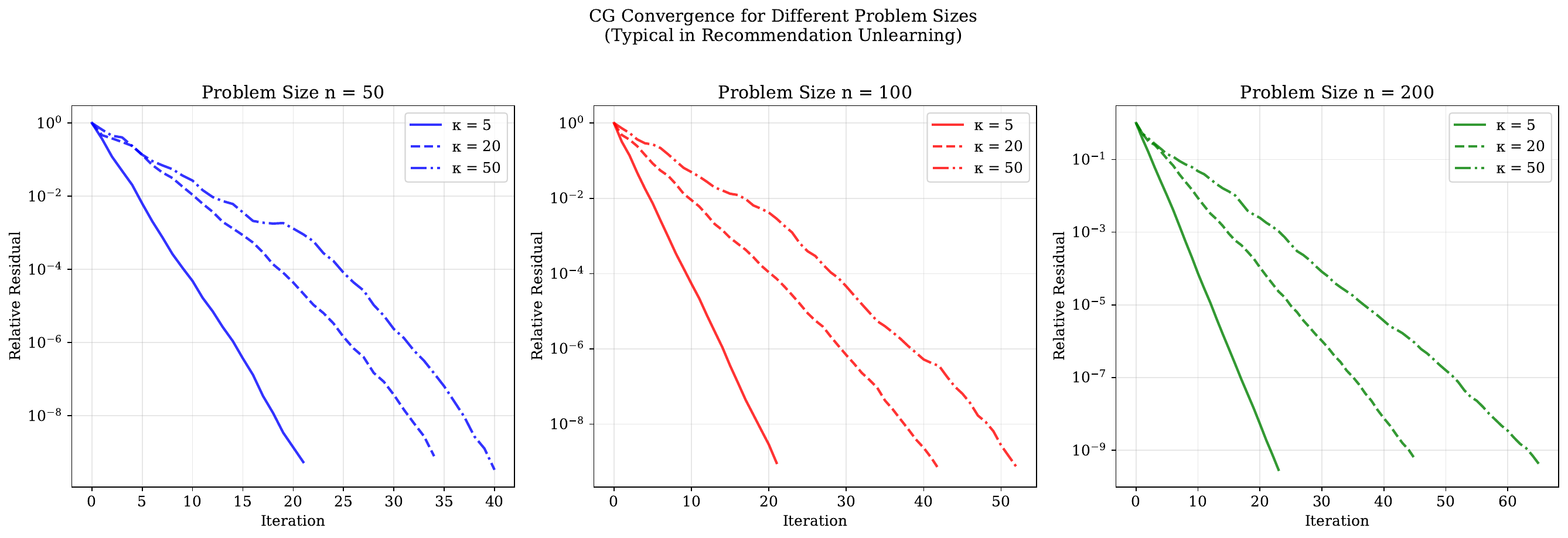}
    \caption{CG convergence behavior for different problem sizes typical in recommendation unlearning. The plots demonstrate consistent convergence patterns across various influenced scope sizes with moderate condition numbers.}
    \label{fig:cg_practical_behavior}
\end{figure}

The practical convergence behavior across different problem sizes, shown in Figure~\ref{fig:cg_practical_behavior}, validates our approach's scalability. Even for larger influenced scopes (n = 200), CG maintains efficient convergence within reasonable iteration counts.

\subsection{Stability Enhancements}

To ensure robust performance across diverse scenarios, our CG implementation incorporates several stability enhancements:

\textbf{Adaptive Tolerance.} The termination tolerance is dynamically adjusted based on problem characteristics and iteration progress to balance computational efficiency with solution accuracy.

\textbf{Iteration Limit.} We impose a maximum iteration limit to prevent excessive computation in pathological cases, typically set to $\min(n, 1000)$ where $n$ is the problem dimension.

\textbf{Numerical Monitoring.} The algorithm monitors key numerical indicators such as residual norm progression and direction orthogonality to detect potential instabilities and trigger corrective actions when necessary.

These convergence and stability properties ensure that the CG algorithm provides reliable and efficient solutions to the linear systems arising in L2UnRank, contributing to the overall robustness and effectiveness of our unlearning framework.

\end{document}